\documentclass[iicol,sn-basic,pdflatex]{sn-jnl}

\newcommand{\bm}[1]{\boldsymbol{#1}}
\newcommand{\tr}[1]{\textrm{tr}\left(#1\right)}
\newcommand{\trans}{\scriptscriptstyle \textrm{T}}
\newcommand{\inv}{\scriptscriptstyle -1}
\newcommand{\Bcal}{\mathcal{B}}
\newcommand{\EDF}{\textsf{edf}}
\newcommand{\REDF}{\textsf{redf}}
\newcommand{\minrho}{\rho_{\min}}
\newcommand{\maxrho}{\rho_{\max}}
\newcommand{\maxrhohat}{\hat{\rho}_{\max}}

\begin{document}

\title[Automatic Search Intervals]{Automatic Search Intervals for the Smoothing Parameter in Penalized Splines}

\author[1]{\fnm{Zheyuan} \sur{Li}}\email{zheyuan.li@vip.henu.edu.cn}

\author*[2]{\fnm{Jiguo} \sur{Cao}}\email{jiguo\_cao@sfu.ca}

\affil[1]{\orgdiv{School of Mathematics and Statistics}, \orgname{Henan University}, \orgaddress{\city{Kaifeng}, \state{Henan}, \country{China}}}

\affil[2]{\orgdiv{Department of Statistics and Actuarial Science}, \orgname{Simon Fraser University}, \orgaddress{\city{Burnaby}, \state{British Columbia}, \country{Canada}}}

\abstract{The selection of smoothing parameter is central to the estimation of penalized splines. The best value of the smoothing parameter is often the one that optimizes a smoothness selection criterion, such as generalized cross-validation error (GCV) and restricted likelihood (REML). To correctly identify the global optimum rather than being trapped in an undesired local optimum, grid search is recommended for optimization. Unfortunately, the grid search method requires a pre-specified search interval that contains the unknown global optimum, yet no guideline is available for providing this interval. As a result, practitioners have to find it by trial and error. To overcome such difficulty, we develop novel algorithms to automatically find this interval. Our automatic search interval has four advantages. (i) It specifies a smoothing parameter range where the associated penalized least squares problem is numerically solvable. (ii) It is criterion-independent so that different criteria, such as GCV and REML, can be explored on the same parameter range. (iii) It is sufficiently wide to contain the global optimum of any criterion, so that for example, the global minimum of GCV and the global maximum of REML can both be identified. (iv) It is computationally cheap compared with the grid search itself, carrying no extra computational burden in practice. Our method is ready to use through our recently developed \textbf{R} package \textbf{gps} (\textgreater= version 1.1). It may be embedded in more advanced statistical modeling methods that rely on penalized splines. \kern 10.0cm}


\maketitle

\section{Introduction}

Penalized splines are flexible and popular tools for estimating unknown smooth functions. They have been applied in many statistical modeling frameworks, including generalized additive models \citep{WOS:000457464800009}, single-index models \citep{WOS:000440611000006}, generalized partially linear single-index models \citep{WOS:000395004300017}, functional mixed-effects models \citep{WOS:000444443000004}, survival models \citep{WOS:000641387600001,WOS:000364259800022}, trajectory modeling for longitudinal data \citep{WOS:000418746100004,WOS:000436403600033}, additive quantile regression models \citep{WOS:000549910800001}, varying coefficient models \citep{WOS:000450660500010}, quantile varying coefficient models \citep{WOS:000434068100003}, spatial models \citep{WOS:000448217800001,WOS:000426321800004}, spatiotemporal models \citep{WOS:000489511900001,WOS:000456529100005}, spatiotemporal quantile and expectile regression models \citep{WOS:000461592800005,WOS:000546038500003}.

To explain the fundamental idea of a penalized spline, consider the following smoothing model for observations $(x_i, y_i)$, $i = 1, \ldots, n$:
\begin{equation*}
	y_i = f(x_i) + e_i, \kern 5mm e_i \overset{\scriptscriptstyle\textrm{iid}}{\sim} \textrm{N}(0, \sigma^2).
\end{equation*}
After expressing $f(x) = \sum_{j = 1}^{p}\Bcal_j(x)\beta_j$ with some spline basis $\Bcal_j(x)$, $j = 1, \ldots, p$, we estimate basis coefficients $\bm{\beta} = (\beta_1, \beta_2, \ldots, \beta_p)^{\trans}$ by minimizing the penalized least squares (PLS) objective:
\begin{equation}
	\label{eqn: PLS}
	\|\bm{y} - \bm{B\beta}\|^2 + \textrm{e}^{\rho}\bm{\beta}^{\trans}\bm{S}\bm{\beta}.
\end{equation}
The first term is the least squares, where $\bm{y} = (y_1, y_2, \ldots, y_n)^{\trans}$ and $\bm{B}$ is a design matrix whose $(i,j)$\textsuperscript{th} entry is $\Bcal_j(x_i)$. The second term is the penalty, where $\bm{S}$ is a penalty matrix (such that $\bm{\beta}^{\trans}\bm{S}\bm{\beta}$ is a wiggliness measure for $f$) and $\rho \in (-\infty, +\infty)$ is a smoothing parameter controlling the strength of the penalization. The choice of $\rho$ is critical, as it trades off $f$'s closeness to data for $f$'s smoothness. The best $\rho$ value often optimizes a smoothness selection criterion, like generalized cross-validation error (GCV) \citep{Wahba-spline-models-book} and restricted likelihood (REML) \citep{Wood-GAMs-book}. Many strategies can be applied to this optimization task. However, to correctly identify the global optimum rather than being trapped in a local optimum, grid search is recommended. Specifically, we attempt an equidistant grid of $\rho$ values in a search interval $[\minrho, \maxrho]$ and pick the one that minimizes GCV or maximizes REML.

\begin{figure}
	\includegraphics[width = \columnwidth]{"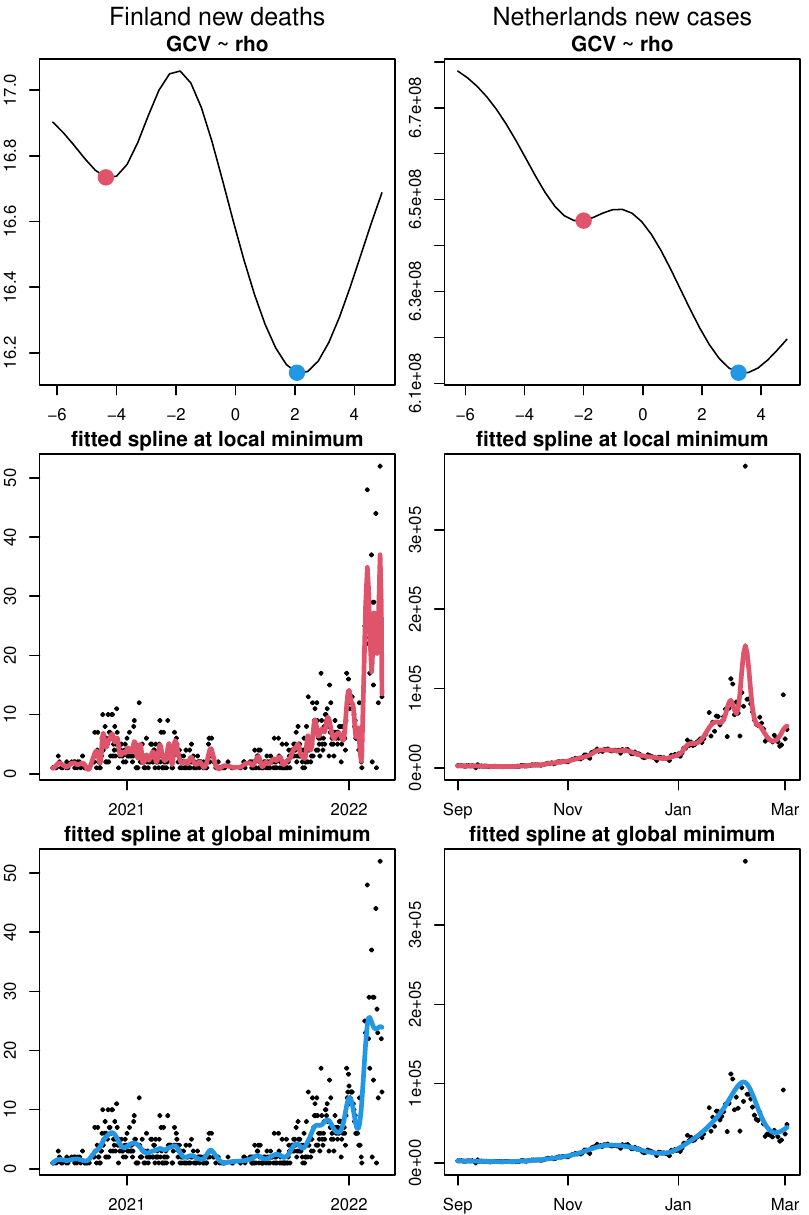"}
	\caption{Daily COVID-19 data smoothing. Left column: new deaths in Finland from 2020-09-01 to 2022-03-01; Right column: new cases in Netherlands from 2021-09-01 to 2022-03-01. The first row shows that GCV has a local minimum (red) and a global minimum (blue). The second row shows that the fitted spline corresponding to the local minimum is wiggly. The last row shows that the fit corresponding to the global minimum is smoother and more plausible.}
	\label{fig1}
\end{figure}

To illustrate that a criterion can have multiple local optima, and mistaking a local optimum for the global optimum may give undesirable result, consider smoothing daily new deaths attributed to COVID-19 in Finland from 2020-09-01 to 2022-03-01 and daily new confirmed cases of COVID-19 in Netherlands from 2021-09-01 to 2022-03-01 (data source: Our World in Data \citep{owidcoronavirus}). Figure \ref{fig1} shows that the GCV (against $\rho$) in each example has a local minimum and a global minimum. The fitted spline corresponding to the local minimum is very wiggly, especially for Finland. Instead, the fit corresponding to the global minimum is smoother and more plausible; for example, it reasonably depicts the single peak during the first Omicron wave in Netherlands. In both cases, minimizing GCV via gradient descent or Newton's method can be trapped in the local minimum, if the initial guess for $\rho$ is in its neighborhood. By contrast, doing a grid search in $[-6, 5]$ guarantees that the global minimum can be found.

In general, to successfully identify the global optimum of a criterion function in grid search, the search interval $[\minrho, \maxrho]$ must be sufficiently wide so that the criterion can be fully explored. Unfortunately, no guideline is available for pre-specifying this interval, so practitioners have to find it by trial and error. This inevitably causes two problems. First of all, the PLS problem \eqref{eqn: PLS} is not numerically solvable  when $\rho$ is too large, because the limiting linear system for $\bm{\beta}$ is rank-deficient. A $\rho$ value too small triggers the same problem if $p > n$, because the limiting unpenalized regression spline problem has no unique solution. As a result, practitioners do not even know what $\rho$ range is numerically safe to attempt. Secondly, a PLS solver is often a computational kernel for more advanced smoothing problems, like robust smoothing with M-estimators \citep{DreassiEmanuela2014SMrf,WOS:000374563100005} and  backfitting-based generalized additive models \citep{WOS:000179809600002,WOS:000646481900001,WOS:000372035600006}. In these problems, a PLS needs be solved at each iteration for reweighted data, and it is more difficult to pre-specify a search interval because it changes from iteration to iteration.

To address these issues, we develop algorithms that automatically find a suitable $[\minrho, \maxrho]$ for the grid search. Our automatic search interval has four advantages. (i) It gives a safe $\rho$ range where the PLS problem \eqref{eqn: PLS} is numerically solvable. (ii) It is criterion-independent so that different smoothness selection criteria, including GCV and REML, can be explored on the same $\rho$ range. (iii) It is wide enough to contain the global optimum of any selection criterion, such as the global minimum of GCV and the global maximum of REML. (iv) It is computationally cheap compared with the grid search itself, carrying no extra cost in practice.

We will introduce our method using penalized B-splines, because it is particularly challenging to develop algorithms for this class of penalized splines to achieve (iv). We will describe how to adapt our method for other types of penalized splines at the end of this paper. Our method is ready to use via our \textbf{R} package \textbf{gps} ($\ge$ version 1.1) \citep{gps-package}. It may be embedded in more advanced statistical modeling methods that rely on penalized splines.

The rest of this paper is organized as follows: Section \ref{section: penalized B-splines} introduces penalized B-splines and their estimation for any trial $\rho$ value. Section \ref{section: automatic search intervals} derives, discusses and illustrates several search intervals for $\rho$. Section \ref{section: Simulations for heuristic upper bound} conducts simulations for these intervals. Section \ref{section: application} applies our search interval to practical smoothing by revisiting the COVID-19 example, with a comparison between GCV and REML in smoothness selection. The last section summarizes our method and addresses reviewers' concerns. Our \textbf{R} code is available on the internet at \url{https://github.com/ZheyuanLi/gps-vignettes/blob/main/gps2.pdf}.

\section{Penalized B-Splines}
\label{section: penalized B-splines}

There exists a wide variety of penalized splines, depending on the setup of basis and penalty. If we choose the basis functions $\Bcal_j(x)$, $j = 1, \ldots, p$ to be B-splines \citep{deBoor-book} and accompany them by either a difference penalty or a derivative penalty, we arrive at a particular class of penalized splines: the penalized B-splines family. This family has three members: O-splines \citep{O-splines-1}, standard P-splines \citep{P-splines} and general P-splines \citep{gps-paper}; see the last reference for a good overview of this family. For this family, the matrix $\bm{S}$ in \eqref{eqn: PLS} can be induced by $\bm{S} = \bm{D}_m^{\trans}\bm{D}_m$. Therefore, the PLS objective can be expressed as:
\begin{equation}
	\label{eqn: penalized B-splines}
	\|\bm{y} - \bm{B\beta}\|^2 + \textrm{e}^{\rho}\|\bm{D}_m\bm{\beta}\|^2.
\end{equation}
In this context, $\bm{B}$ is an $n \times p$ design matrix for B-splines of order $d \ge 2$, and $\bm{D}_m$ is a $(p - m) \times p$ penalty matrix of order $1 \le m \le d - 1$. Basically, the three family members only differ in (i) the type of knots for constructing B-splines; (ii) the penalty matrix $\bm{D}_m$ applied to B-spline coefficients $\bm{\beta}$. See Appendix \ref{Appendix A} for concrete examples.

When estimating penalized B-splines, we can pre-compute some $\rho$-independent quantities that are reused over $\rho$ values on a search grid. These include $\|\bm{y}\|^2$, $\bm{B^{\trans}B}$, $\bm{D}_m^{\trans}\bm{D}_m$, $\bm{B^{\trans}y}$ and the lower triangular factor $\bm{L}$ in the Cholesky factorization $\bm{B^{\trans}B} = \bm{LL^{\trans}}$. In the following, we will consider them to be known quantities so that they do not count toward the overall computational cost of the estimation procedure.

Now, for any $\rho$ value on a search grid, the PLS solution to \eqref{eqn: penalized B-splines} is $\bm{\hat{\beta}} = \bm{C}^{\inv}\bm{B^{\trans}y}$, where $\bm{C} = \bm{B^{\trans}B} + \textrm{e}^{\rho}\bm{D}_m^{\trans}\bm{D}_m$. The vector of fitted values is $\bm{\hat{y}} = \bm{B\hat{\beta}} = \bm{B}\bm{C}^{\inv}\bm{B^{\trans}y}$. By definition, the sum of squared residuals is $\textrm{RSS}(\rho) = \|\bm{y} - \bm{\hat{y}}\|^2$, but it can also be computed without using $\bm{\hat{y}}$:
\begin{equation*}
	\begin{split}
		\textrm{RSS}(\rho) &= \|\bm{y} - \bm{B\hat{\beta}}\|^2
		= (\bm{y} - \bm{B\hat{\beta}})^{\trans}(\bm{y} - \bm{B\hat{\beta}})\\
		&= \bm{y^{\trans}y} - 2\bm{\hat{\beta}^{\trans}B^{\trans}y} + \bm{\hat{\beta}^{\trans}B^{\trans}B\hat{\beta}}\\
		&= \bm{y^{\trans}y} - 2\bm{\hat{\beta}^{\trans}B^{\trans}y} + \bm{\hat{\beta}^{\trans}LL^{\trans}\hat{\beta}}\\
		&= \|\bm{y}\|^2 - 2\bm{\hat{\beta}^{\trans}B^{\trans}y} + \|\bm{L^{\trans}\hat{\beta}}\|^2.
	\end{split}
\end{equation*}
The complexity of the fitted spline is measured by its effective degree of freedom (\EDF), defined as the trace of the hat matrix mapping $\bm{y}$ to $\bm{\hat{y}}$:
\begin{equation*}
	\EDF(\rho) = \tr{\bm{B}\bm{C}^{\inv}\bm{B^{\trans}}} = \tr{\bm{C}^{\inv}\bm{B^{\trans}B}}.
\end{equation*}
The GCV error can then be calculated using RSS and \EDF:
\begin{equation*}
	\textrm{GCV}(\rho) = \frac{n\cdot\textrm{RSS}}{(n - \EDF)^2}.
\end{equation*}
The restricted likelihood $\textrm{REML}(\rho)$ has a more complicated from and is detailed in Appendix \ref{Appendix B}. All these quantities vary with $\rho$, explaining why the choice of $\rho$ affects the resulting fit. It should also be pointed out that although {\EDF} depends on the basis $\bm{B}$ and the penalty $\bm{D}_m$, it depends on neither the response data $\bm{y}$ nor the choice of the smoothness selection criterion.

The actual computations of $\bm{\hat{\beta}}$ and {\EDF} do not require explicitly forming $\bm{C}^{\inv}$. We can compute the Cholesky factorization $\bm{C} = \bm{KK^{\trans}}$ for the lower triangular factor $\bm{K}$, then solve triangular systems $\bm{Kx} = \bm{B^{\trans}y}$ and $\bm{K^{\trans}\hat{\beta}} = \bm{x}$ for $\bm{\hat{\beta}}$. The {\EDF} can also be computed using the Cholesky factors:
\begin{equation}
	\label{eqn: edf-via-pls}
	\begin{split}
		\EDF(\rho)
		&= \tr{\bm{C}^{\inv}\bm{LL^{\trans}}}
		= \tr{\bm{L^{\trans}}\bm{C}^{\inv}\bm{L}}\\
		&= \textrm{tr}\big((\bm{K}^{\inv}\bm{L})^{\trans}\bm{K}^{\inv}\bm{L}\big)
		= \|\bm{K}^{\inv}\bm{L}\|_{\scriptscriptstyle F}^2,
	\end{split}
\end{equation}
where we solve the triangular system $\bm{KX} = \bm{L}$ for $\bm{X} = \bm{K}^{\inv}\bm{L}$, and calculate the squared Frobenius norm $\|\bm{X}\|_{\scriptscriptstyle F}^2$ (the sum of squared elements in $\bm{X}$).

Penalized B-splines are sparse and computationally efficient. Notably, $\bm{B^{\trans}B}$, $\bm{D}_m^{\trans}\bm{D}_m$, $\bm{C}$, $\bm{L}$ and $\bm{K}$ are band matrices, so the aforementioned Cholesky factorizations and triangular systems can be computed and solved in $O(p^2)$ floating point operations. That is, both $\bm{\hat{\beta}}$ and {\EDF} come at $O(p^2)$ cost. The computation of RSS (without using $\bm{\hat{y}}$) has $O(p)$ cost. Once RSS and {\EDF} are known, GCV is known. In addition, Appendix \ref{Appendix B} shows that the REML score can be obtained at $O(p)$ cost. Thus, computations for a given $\rho$ value have $O(p^2)$ cost. When $\rho$ is selected over $N$ trial values on a search grid to optimize GCV or REML, the overall computational cost is $O(Np^2)$.

\section{Automatic Search Intervals}
\label{section: automatic search intervals}

To start with, we formulate {\EDF} in a different way that better reveals its mathematical property. Let $q = p - m$. We form the $p \times q$ matrix (at $O(p^2)$ cost):
\begin{equation}
	\label{eqn: matrix E}
	\bm{E} = \bm{L}^{\inv}\bm{D}_m^{\trans},
\end{equation}
and express $\bm{C}$ as:
\begin{equation*}
	\bm{C} = \bm{LL^{\trans}} + \textrm{e}^{\rho}\bm{D}_m^{\trans}\bm{D}_m = \bm{L}[\bm{I} + \textrm{e}^{\rho}\bm{EE^{\trans}}]\bm{L^{\trans}}\\
\end{equation*}
Plugging this into $\EDF(\rho) = \tr{\bm{L^{\trans}}\bm{C}^{\inv}\bm{L}}$ (see \eqref{eqn: edf-via-pls}), we get:
\begin{equation*}
	\EDF(\rho) = \textrm{tr}\big([\bm{I} + \textrm{e}^{\rho}\bm{EE^{\trans}}]^{\inv}\big),
\end{equation*}
where $\bm{I}$ is an identity matrix. The $p \times p$ symmetric matrix $\bm{EE^{\trans}}$ is positive semi-definite with rank $q$. Thus, it has $q$ positive eigenvalues $\lambda_1 > \lambda_2 > \ldots > \lambda_q$, followed by $m$ zero eigenvalues. Let its eigendecomposition be $\bm{EE^{\trans}} = \bm{U\Lambda U}^{\inv}$, where $\bm{U}$ is an orthonormal matrix with eigenvectors and $\bm{\Lambda}$ is a $p \times p$ diagonal matrix with eigenvalues. Then,
\begin{equation*}
	\begin{split}
		\EDF(\rho) &= \textrm{tr}\big([\bm{U}\bm{U}^{\inv} + \textrm{e}^{\rho}\bm{U\Lambda U}^{\inv}]^{\inv}\big)\\
		&= \textrm{tr}\big([\bm{U}(\bm{I} + \textrm{e}^{\rho}\bm{\Lambda})\bm{U}^{\inv}]^{\inv}\big)\\
		&= \textrm{tr}\big(\bm{U}(\bm{I} + \textrm{e}^{\rho}\bm{\Lambda})^{\inv}\bm{U}^{\inv}\big)\\
		&= \textrm{tr}\big((\bm{I} + \textrm{e}^{\rho}\bm{\Lambda})^{\inv}\big)\\
		&= m + \textstyle\sum_{j = 1}^{q}(1 + \textrm{e}^{\rho} \lambda_j)^{\inv}.
	\end{split}
\end{equation*}
We call $(\lambda_j)_1^{q}$ the Demmler-Reinsch eigenvalues, in tribute to \citet{eigen-analysis-of-periodic-smoothing-spline} who first studied the eigenvalue problem of smoothing splines \citep{Reinsch-cubic-smoothing-spline,Reinsch-smoothing-spline,Yuedong2011}. We also define the restricted {\EDF} as:
\begin{equation}
	\label{eqn: redf}
	\REDF(\rho) = \EDF(\rho) - m = \sum_{j = 1}^{q}\frac{1}{1 + \textrm{e}^{\rho} \lambda_j},
\end{equation}
which is a monotonically decreasing function of $\rho$. As $\rho \to -\infty$, it increases to $q$; as $\rho \to +\infty$, it decreases to 0. Given this one-to-one correspondence between $\rho$ and \REDF, it is clear that choosing the optimal $\rho$ is equivalent to choosing the optimal \REDF.

\subsection{An Exact Search Interval}

The relation between $\rho$ and {\REDF} is enlightening. Although it is difficult to see what $[\minrho,\maxrho]$ is wide enough for searching for $\rho$, it is easy to see what $[\REDF_{\min},\REDF_{\max}]$ is adequate for searching for \REDF. For example,
\begin{equation}
	\label{eqn: redf bounds}
	\begin{split}
	\REDF_{\min} &= q\kappa,\\
	\REDF_{\max} &= q(1 - \kappa),
	\end{split}
\end{equation}
with a small $\kappa$ are reasonable. We interpret
$\kappa$ as a coverage parameter, as $[\REDF_{\min},\REDF_{\max}]$ covers $100(1 - 2\kappa)$\% of $[0, q]$.  In practice, $\kappa = 0.01$ is good enough, in which case $[\REDF_{\min},\REDF_{\max}]$ covers 98\% of $[0, q]$. It then follows that $\minrho$ and $\maxrho$ satisfy
\begin{equation}
	\label{eqn: exact rho bounds}
	\begin{split}
	\REDF(\minrho) &= \REDF_{\max},\\
	\REDF(\maxrho) &= \REDF_{\min},
	\end{split}
\end{equation}
and can be solved for via root-finding. As $\REDF(\rho)$ is differentiable, we can use Newton's method. See Algorithm \ref{algorithm: root-finding} for an implementation of this method for a general root-finding problem $g(x) = 0$.

\begin{algorithm}
	\caption{Newton's method for finding the root of $g(x) = 0$. Inputs: (i) $x$, an initial value, (ii) $\delta_{\max}$, maximum size of a Newton step.}
	\label{algorithm: root-finding}
	\begin{algorithmic}[1]
		\State $g = g(x)$
		\Loop
		\State $g' = g'(x)$
		\State $\delta = -g'/g$
		\If {$\lvert\delta\rvert < \lvert g\rvert10^{-6}$}
		\State break
		\EndIf
		\State $\delta = \textrm{sign}(\delta)\cdot\min(\lvert\delta\rvert, \delta_{\max})$
		\Loop
		\State $\tilde{x} = x + \delta_k$
		\State $\tilde{g} = g(\tilde{x})$
		\If {$\lvert \tilde{g}\rvert < \lvert g\rvert$}
		\State break
		\EndIf
		\State $\delta = \delta / 2$
		\EndLoop
		\State $x = \tilde{x}$
		\State $g = \tilde{g}$
		\EndLoop\\
		\Return $x$
	\end{algorithmic}
\end{algorithm}

It should be pointed out that Algorithm \ref{algorithm: root-finding} is not fully automatic. To make it work, we need an initial $\rho$ value, as well as a suitable $\delta_{\max}$ for bounding the size of a Newton step (which helps prevent overshooting). This is not easy, though, since we don't know a sensible range for $\rho$ (in fact, we are trying to find such a range). We will come back to this issue in the next section. For now, let's discuss the strength and weakness of this approach.

The interval $[\minrho, \maxrho]$ is back-transformed from \REDF. As is stressed in Section \ref{section: penalized B-splines}, {\EDF} (and thus \REDF) does not depend on $y$ values or the choice of the smoothness selection criterion; neither does the interval. This is a nice property. It is also an advantage of our method, for there is no need to find a new interval when $y$ values or the criterion change. We call this idea of back-transformation the \REDF-oriented thinking.

Nevertheless, the interval is computationally expensive. To get $(\lambda_j)_1^{q}$, an eigendecomposition at $O(p^3)$ cost is needed. Section \ref{section: penalized B-splines} shows that solving PLS along with grid search only has $O(Np^2)$ cost. Thus, when $p$ is big, finding the interval is even more costly than the subsequent grid search. This is unacceptable and we need a better strategy.

\subsection{A Wider Search Interval}

In fact, we don't have to find the exact $[\minrho, \maxrho]$ that satisfies \eqref{eqn: redf bounds} and \eqref{eqn: exact rho bounds}. It suffices to find a wider interval:
\begin{equation*}
	[\minrho^*, \maxrho^*] \supseteq [\minrho, \maxrho].
\end{equation*}
That is, find $\minrho^*$ and $\maxrho^*$ such that $\minrho \ge \minrho^*$ and $\maxrho \le \maxrho^*$. Surprisingly, in this way, we only need the maximum and the minimum eigenvalues ($\lambda_1$ and $\lambda_q$) instead of all the eigenvalues.

We now derive this wider interval. From $\lambda_1 \geq \lambda_j \geq \lambda_q$, we have:
\begin{equation*}
	\frac{1}{1 + \textrm{e}^{\rho} \lambda_1} \leq \frac{1}{1 + \textrm{e}^{\rho} \lambda_j} \leq \frac{1}{1 + \textrm{e}^{\rho} \lambda_q}.
\end{equation*}
Then, applying $\sum_{j = 1}^{q}$ to these terms, we get:
\begin{equation*}
	\frac{q}{1 + \textrm{e}^{\rho} \lambda_1} \leq \REDF(\rho) \leq \frac{q}{1 + \textrm{e}^{\rho} \lambda_q}.
\end{equation*}
Since the result holds for any $\rho$, including $\minrho$ and $\maxrho$, there are:
\begin{equation*}
	\begin{split}
		\REDF(\minrho) &\geq \frac{q}{1 + \textrm{e}^{\minrho}\lambda_1},\\
		\REDF(\maxrho) &\leq \frac{q}{1 + \textrm{e}^{\maxrho}\lambda_q}.
	\end{split}
\end{equation*}
Together with \eqref{eqn: redf bounds} and \eqref{eqn: exact rho bounds}, we see:
\begin{equation*}
	\begin{split}
		q(1 - \kappa) &\geq \frac{q}{1 + \textrm{e}^{\minrho}\lambda_1},\\
		q\kappa &\leq \frac{q}{1 + \textrm{e}^{\maxrho}\lambda_q}.
	\end{split}
\end{equation*}
These imply:
\begin{equation*}
	\begin{split}
		\minrho &\ge \log\left(\frac{\kappa}{(1 - \kappa)\lambda_1}\right) = \minrho^*,\\
		\maxrho &\le \log\left(\frac{1 - \kappa}{\kappa \lambda_q}\right) = \maxrho^*.
	\end{split}
\end{equation*}

Better yet, we can replace the maximum eigenvalue $\lambda_1$ by the mean eigenvalue $\bar{\lambda} = \sum_{j = 1}^q\lambda_j/q$ for a tighter lower bound. In general, the harmonic mean of positive numbers $(a_j)_1^q$ is no larger than their arithmetic mean:
\begin{equation*}
	\frac{q}{\sum_{j = 1}^q\frac{1}{a_j}} \leq \bar{a} \kern 2mm \Rightarrow \kern 2mm \frac{q}{\bar{a}} \leq \sum_{j = 1}^q\frac{1}{a_j}.
\end{equation*}
If we set $a_j = 1 + \textrm{e}^{\rho} \lambda_j$, we get:
\begin{equation*}
	\frac{q}{1 + \textrm{e}^{\rho}\bar{\lambda}} \leq \REDF(\rho).
\end{equation*}
This allows us to update $\minrho^*$. In the end, we have:
\begin{equation}
	\label{eqn: rho bounds}
	\begin{split}
		\minrho^* &= \log\left(\frac{\kappa}{(1 - \kappa)\bar{\lambda}}\right),\\
		\maxrho^* &= \log\left(\frac{1 - \kappa}{\kappa \lambda_q}\right).
	\end{split}
\end{equation}

At first glance, it appears that we still need all the eigenvalues in order to compute their mean $\bar{\lambda}$. But the trick is that $(\lambda_j)_1^q$ add up to:
\begin{equation*}
	\sum_{j = 1}^{q}\lambda_j = \tr{\bm{\Lambda}} = \tr{\bm{U\Lambda U}^{\inv}} = \tr{\bm{EE^{\trans}}} = \|\bm{E}\|_{\scriptscriptstyle F}^2,
\end{equation*}
so we can easily compute $\bar{\lambda}$ (at $O(p^2)$ cost):
\begin{equation}
	\label{eqn: MeanEigen}
	\bar{\lambda} = \frac{1}{q}\sum_{j = 1}^{q}\lambda_j = \frac{\|\bm{E}\|_{\scriptscriptstyle F}^2}{q}.
\end{equation}
This leaves the minimum eigenvalue $\lambda_q$ the only nontrivial quantity to compute. As we will see in the next section, the computation of $\lambda_q$ merely involves $O(p^2)$ complexity. Therefore, the wider search interval $[\minrho^*, \maxrho^*]$ is available at $O(p^2)$ cost, much cheaper than the exact search interval $[\minrho, \maxrho]$ that comes at $O(p^3)$ cost.

Another advantage of the wider interval is its closed-form formula. The exact interval is defined implicitly by root-finding, and its computation is not automatic as Algorithm \ref{algorithm: root-finding} requires $\rho$-relevant inputs. However, the wider interval can be directly computed using eigenvalues that are irrelevant to $\rho$, so its computation is fully automatic.

In practice, we can utilize the wider interval to automate the computation of the exact interval. For example, when applying Algorithm \ref{algorithm: root-finding} to find $\minrho$ or $\maxrho$, we may use $(\minrho^* + \maxrho^*)/2$ for its initial value, and bound the size of a Newton step by $\delta_{\max} = (\maxrho^* - \minrho^*)/4$. This improvement does not by any means make the computation of the exact interval less expensive, though.

\subsection{Computational Details}

We now describe $O(p^2)$ algorithms for computing the maximum and the minimum eigenvalues, i.e., $\lambda_1$ and $\lambda_q$, to aid the fast computation of the wider interval $[\minrho^*, \maxrho^*]$. Note that although $\lambda_1$ does not show up in \eqref{eqn: rho bounds}, it is useful for assessing the credibility of the computed $\lambda_q$, which will soon be explained.

Recall that $(\lambda_j)_1^q$ are the positive eigenvalues of the $p \times p$ positive semi-definite matrix $\bm{EE^{\trans}}$, and the matrix also has $m$ zero eigenvalues. To get rid of these nuisance eigenvalues, we can work with the $q \times q$ positive-definite matrix $\bm{E^{\trans}E}$ instead. Here, the trick is that $\bm{E^{\trans}E}$ and $\bm{EE^{\trans}}$ have the same positive eigenvalues that equal the squared nonzero singular values of $\bm{E}$. This can be proved using the singular value decomposition of $\bm{E}$.

In general, given a positive definite matrix $\bm{A}$, we compute its maximum eigenvalue using power iteration and its minimum eigenvalue using inverse iteration. The two algorithms iteratively compute $\bm{Av}$ and $\bm{A^{\inv}v}$, respectively. Unfortunately, when $\bm{A} = \bm{E^{\trans}E}$, the latter operation is as expensive as $O(p^3)$ because $\bm{A}$ is fully dense. Therefore, naively applying inverse iteration to compute $\lambda_q$ is not any cheaper than a full eigendecomposition.

To obtain $\lambda_q$ at $O(p^2)$ cost, we are to exploit the following partitioning:
\begin{equation*}
	\bm{E} = \begin{bmatrix} \bm{E_1} \\ \bm{E_2} \end{bmatrix},
\end{equation*}
where $\bm{E_1}$ is a $q \times q$ lower triangular matrix and $\bm{E_2}$ is an $m \times q$ rectangular matrix. Below is an illustration of such structure (with $p = 6$, $m = 2$ and $q = 4$), where `$\times$' and `$\circ$' denote the nonzero elements in $\bm{E_1}$ and $\bm{E_2}$, respectively.
\begin{equation*}
	\left[\begin{array}{cccc}
		\times\\
		\times & \times\\
		\times & \times & \times\\
		\times & \times & \times & \times\\
		\hline
		\circ & \circ & \circ & \circ\\
		\circ & \circ & \circ & \circ
	\end{array}\right]
\end{equation*}
Such ``trapezoidal'' structure exclusively holds for the penalized B-splines family, and it allows us to express $\bm{A}$ as an update to $\bm{E_1^{\trans}E_1}$:
\begin{equation*}
	\bm{A} = \bm{E^{\trans}E} = \bm{E_1^{\trans}E_1} + \bm{E_2^{\trans}E_2}.
\end{equation*}
Using Woodbury identity \citep{woodbury}, we obtain an explicit inversion formula:
\begin{equation*}
	\begin{split}
	\bm{A}^{\inv} &= (\bm{E_1^{\trans}E_1})^{\inv} - \bm{F}(\bm{I} + \bm{R^{\trans}R})^{\inv}\bm{F^{\trans}}\\
	&= (\bm{E_1^{\trans}E_1})^{\inv} - \bm{F}(\bm{GG^{\trans}})^{\inv}\bm{F^{\trans}},
	\end{split}
\end{equation*}
where $\bm{R} = (\bm{E_1^{\trans}})^{\inv}\bm{E_2^{\trans}}$ and $\bm{F} = \bm{E_1^{\inv}}\bm{R}$ are both $q \times m$ matrices, and $\bm{G}$ is the lower triangular Cholesky factor of the $m \times m$ matrix $\bm{I} + \bm{R^{\trans}R}$. We thus convert the expensive computation of $\bm{A}^{\inv}\bm{v}$ to solving triangular linear systems involving $\bm{E_1}$, $\bm{E_1^{\trans}}$, $\bm{G}$ and $\bm{G^{\trans}}$, which is much more efficient. See Algorithm \ref{algorithm: MinEigen} for implementation details as well as a breakdown of computational costs. The penalty order $m$ is often very small (usually 1, 2 or 3) and the number of iterations till convergence is much smaller than $q$, so the overall cost remains $O(q^2)$ in practice. Since $q = p - m \approx p$, we report this complexity as $O(p^2)$.

\begin{algorithm}
	\caption{Compute the minimum eigenvalue $\lambda_q$ of $\bm{A} = \bm{E^{\trans}E}$. Lines 1 to 5 compute matrices for applying Woodbury identity. Line 6 starts inverse iteration. Lines 10 to 16 compute $\bm{u} = \bm{A}^{\inv}\bm{v}$.}
	\label{algorithm: MinEigen}
	\begin{algorithmic}[1]
		\State $\bm{E} = \left[\begin{smallmatrix}\bm{E_1}\\ \bm{E_2}\end{smallmatrix}\right]$
		\State solve $\bm{E_1^{\trans}R} = \bm{E_2^{\trans}}$ for $\bm{R}$ \Comment{$O(mq^2)$}
		\State solve $\bm{E_1F} = \bm{R}$ for $\bm{F}$ \Comment{$O(mq^2)$}
		\State $\bm{H} = \bm{I} + \bm{R^{\trans}R}$ \Comment{$O(m^3)$}
		\State Cholesky factorization $\bm{H} = \bm{GG^{\trans}}$ \Comment{$O(m^3)$}
		\State initialize $\bm{u}$ as a random vector \Comment{$O(q)$}
		\State $\lambda = 0$
		\Loop
		\State $\bm{v} = \bm{u} / \|\bm{u}\|$ \Comment{$O(q)$}
		\State solve $\bm{E_1^{\trans}a_1} = \bm{v}$ for $\bm{a_1}$ \Comment{$O(q^2)$}
		\State solve $\bm{E_1a_2} = \bm{a_1}$ for $\bm{a_2}$ \Comment{$O(q^2)$}
		\State $\bm{c_1} = \bm{F^{\trans}v}$ \Comment{$O(qm)$}
		\State solve $\bm{Gb_1} = \bm{c_1}$ for $\bm{b_1}$ \Comment{$O(m^2)$}
		\State solve $\bm{G^{\trans}b_2} = \bm{b_1}$ for $\bm{b_2}$ \Comment{$O(m^2)$}
		\State $\bm{c_2} = \bm{Fb_2}$ \Comment{$O(qm)$}
		\State $\bm{u} = \bm{a_2} - \bm{c_2}$ \Comment{$O(q)$}
		\State $\tilde{\lambda} = \bm{v^{\trans}u}$ \Comment{$O(q)$}
		\If {$\tilde{\lambda} < 0$}
		\State Warning: $\bm{E^{\trans}E}$ is numerically singular!
		\State $\lambda_q = 0$
		\State break
		\EndIf
		\If {$\lvert \tilde{\lambda} - \lambda\rvert < \lambda10^{-6}$}
		\State break
		\EndIf
		\State $\lambda = \tilde{\lambda}$
		\EndLoop
		\State $\lambda_q = 1 / \lambda$
		\If {$\lambda_q < \lambda_1\varepsilon$}
		\State Warning: $\bm{E^{\trans}E}$ is numerically singular!
		\State $\lambda_q = \lambda_1\varepsilon$
		\EndIf\\
		\Return $\lambda_q$
	\end{algorithmic}
\end{algorithm}

An important technical detail in Algorithm \ref{algorithm: MinEigen} is that it checks the credibility of the computed $\lambda_q$. Although $\bm{E^{\trans}E}$ is positive definite, in finite precision arithmetic performed by our computers, it becomes numerically singular if $\lambda_q / \lambda_1$ is smaller than the machine precision $\varepsilon$ (the largest positive number such that $1 + \varepsilon = 1$). On modern 64-bit CPUs, this precision is about $1.11 \times 10^{-16}$. In case of numerical singularity, $\lambda_q$ can not be accurately computed for the loss of significant digits, and the output $\lambda_q$ is fake. The best bet in this case, is to reset the computed $\lambda_q$ to $\lambda_1\varepsilon$ (see lines 29 to 31). Sometimes, the computed $\tilde{\lambda}$ at line 17 is negative. This is an early sign of singularity, and we can immediately stop the iteration (see lines 18 to 22). These procedures are necessary safety measures, without which the computed $\lambda_q$ will be too small and the derived $\maxrho^*$ will be too big, making the PLS problem \eqref{eqn: penalized B-splines} unsolvable. In short, Algorithm \ref{algorithm: MinEigen} helps determine a search interval $[\minrho^*, \maxrho^*]$ that is both sufficiently wide and numerically safe for grid search.

We need the maximum eigenvalue $\lambda_1$ before applying Algorithm \ref{algorithm: MinEigen}. The computation of $\lambda_1$ is less challenging: even a direct application of power iteration by iteratively computing $\bm{E^{\trans}}(\bm{Ev})$ is good enough at $O(p^2)$ cost. But we can make it more efficient by exploiting the band sparsity behind the ``factor form'' \eqref{eqn: matrix E} of matrix $\bm{E}$. See Algorithm \ref{algorithm: MaxEigen} for details. The overall complexity is $O(p)$ in practice.

\begin{algorithm}
	\caption{Compute the maximum eigenvalue $\lambda_1$ of $\bm{A} = \bm{E^{\trans}E}$, exploiting the band sparsity of $\bm{L}$ and $\bm{D}_m^{\trans}$ in the ``factor form'' \eqref{eqn: matrix E} of matrix $\bm{E}$. Lines 5 to 8 compute $\bm{u} = \bm{Av}$.}
	\label{algorithm: MaxEigen}
	\begin{algorithmic}[1]
		\State initialize $\bm{u}$ as a random vector \Comment{$O(q)$}
		\State $\lambda = 0$
		\Loop
		\State $\bm{v} = \bm{u} / \|\bm{u}\|$ \Comment{$O(q)$}
		\State $\bm{b} = \bm{D}_m^{\trans}\bm{v}$ \Comment{$O(p)$}
		\State solve $\bm{La_1} = \bm{b}$ for $\bm{a_1}$ \Comment{$O(p)$}
		\State solve $\bm{L^{\trans}a_2} = \bm{a_1}$ for $\bm{a_2}$ \Comment{$O(p)$}
		\State $\bm{u} = \bm{D}_m\bm{a_2}$ \Comment{$O(p)$}
		\State $\tilde{\lambda} = \bm{v^{\trans}u}$ \Comment{$O(q)$}
		\If {$\lvert \tilde{\lambda} - \lambda\rvert < \lambda10^{-6}$}
		\State break
		\EndIf
		\State $\lambda = \tilde{\lambda}$
		\EndLoop\\
		\Return $\lambda$
	\end{algorithmic}
\end{algorithm}

\subsection{An Illustration of the Intervals}

\begin{figure}
	\includegraphics[width = \columnwidth]{"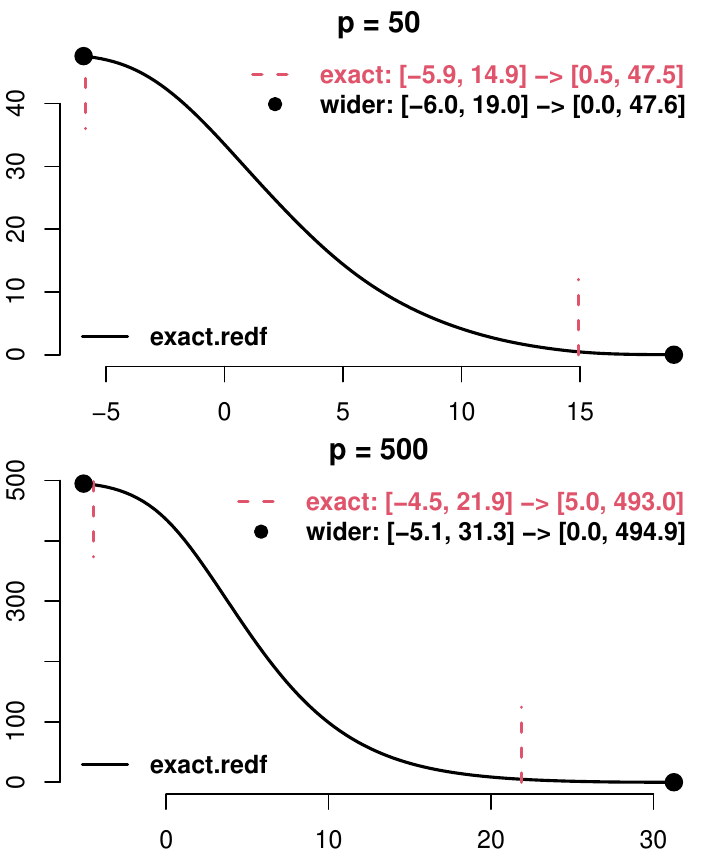"}
	\caption{An example of $\REDF(\rho)$ curve for $p$ cubic B-splines ($d = 4$) on unevenly spaced knots, penalized by a 2nd order ($m = 2$) difference penalty. The black dots ($\minrho^*$ and $\maxrho^*$) lie beyond the red dashed lines ($\minrho$ and $\maxrho$) on both ends, implying that $[\minrho^*, \maxrho^*]$ is wider than $[\minrho, \maxrho]$. The text (with numbers rounded to 1 decimal place) state the actual mapping from $\rho$ range to {\REDF} range.}
	\label{fig2}
\end{figure}

We now illustrate search intervals $[\minrho, \maxrho]$ and $[\minrho^*, \maxrho^*]$ through a simple example. We set up $p$ cubic B-splines ($d = 4$) on unevenly spaced knots and penalized them by a 2nd order ($m = 2$) difference penalty matrix $\bm{D}_2$. We then generated 10 uniformly distributed $x$ values between every two adjacent knots and constructed the design matrix $\bm{B}$ at those locations. Figure \ref{fig2} illustrates the resulting $\REDF(\rho)$ for $p = 50$ and $p = 500$. The nominal mapping from $\rho$ range to {\REDF} range is $(-\infty, +\infty) \to [0, q]$ and here we have $q = p - 2$. For the exact interval, the mapping is $[\minrho, \maxrho] \to [0.01q, 0.99q]$. The wider interval $[\minrho^*, \maxrho^*]$ is mapped to a wider range than $[0.01q, 0.99q]$. 

\begin{figure}
	\includegraphics[width = \columnwidth]{"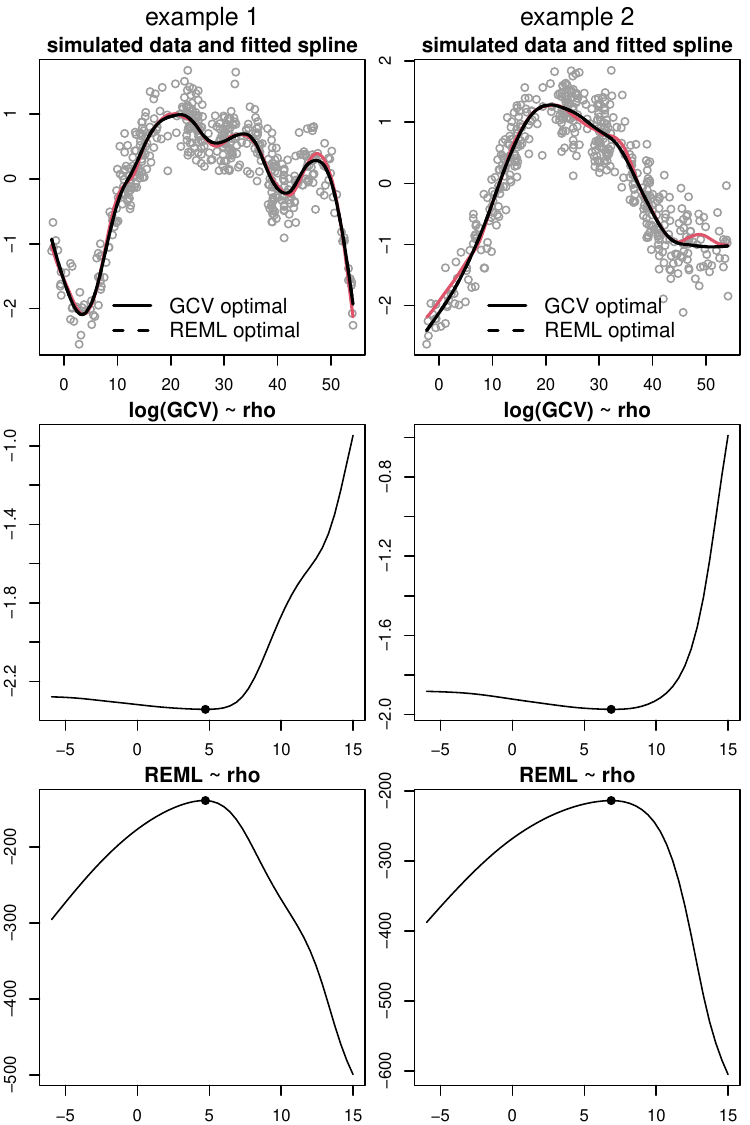"}
	\caption{Two sets of $y$ values are simulated and smoothed for the $p = 50$ case in Figure \ref{fig2}. They yield different GCV or REML curves, but share the same search interval for $\rho$. In the first row, red curves are the true functions and black curves are the optimal spline fit.}
	\label{fig3}
\end{figure}

Our search intervals do not depend on $y$ values or the choice of the smoothness selection criterion. To illustrate this, we simulated two sets of $y$ values for the $p = 50$ case, and chose the optimal $\rho$ by minimizing GCV or maximizing REML. Figure \ref{fig3} shows that while the two datasets yield different GCV or REML curves, they share the same search interval for $\rho$. Incidentally, both GCV and REML choose similar optimal $\rho$ values in each example, leading to indistinguishable fit. This is not always true, though, and we will discuss more about this in Section \ref{section: application}.

\subsection{Heuristic Improvement}

Let's take a second look at Figure \ref{fig2}. As the black circle on the right end is far off the red dashed line, $\maxrho^*$ is a loose upper bound. Moreover, the bigger $p$ is, the looser it is. We now use some heuristics to find a tighter upper bound. Specifically, we seek approximated eigenvalues $(\hat{\lambda}_j)_1^q$ such that $\hat{\lambda}_1 = \lambda_1$, $\hat{\lambda}_q = \lambda_q$ and $\sum_{j = 1}^q\hat{\lambda}_j = q\bar{\lambda}$. Then, replacing $\lambda_j$ by $\hat{\lambda}_j$ in \eqref{eqn: redf} gives an approximated \REDF, with which we can solve \eqref{eqn: redf bounds} and \eqref{eqn: exact rho bounds} for an ``exact'' interval $[\hat{\rho}_{\min}, \maxrhohat]$. We are most interested in $\maxrhohat$. If it is tighter than $\maxrho^*$, even if just empirically, we can narrow the search interval $[\minrho^*, \maxrho^*]$ to $[\minrho^*, \maxrhohat]$. We hereafter call $\maxrhohat$ the heuristic upper bound.

\begin{figure}
	\includegraphics[width = \columnwidth]{"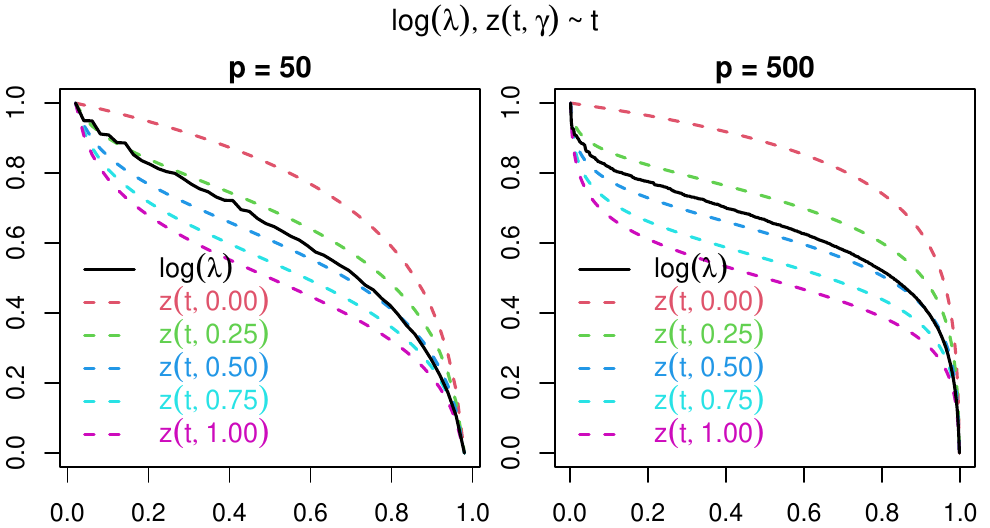"}
	\caption{The asymptotic decay established by Xiao (2019) (red dashed) often does not reflect the actual decay (black solid) of $\log(\lambda_j)$, because the first few eigenvalues may drop faster. Empirically, the actual decay of $\log(\lambda_j)$ resembles an ``S''-shaped curve $z(t_j, \gamma)$, and we can tweak the value of the shape parameter  $\gamma$ to make this curve similar (but not identical) to $\log(\lambda_j)$. (Note: the ranges of $\log(\lambda_j)$ and $z_j$ have been transformed to [0, 1] to aid shape comparison.)}
	\label{fig4}
\end{figure}

For reasonable approximation, knowledge on $\lambda_j$'s decay pattern is useful. Recently, \citet[Lemma 5.1]{Xiao-2019} established an asymptotic decay rate at about $O((1 - \frac{j}{q + m})^{2m}) \sim O((1 - \frac{j + m}{q + m})^{2m})$ (Xiao arranged $(\lambda_j)_1^q$ in ascending order and his original result is $O((\frac{j}{q + m})^{2m}) \sim O((\frac{j + m}{q + m})^{2m})$). So, $\log(\lambda_j)$ roughly decays like $z_j = \log(1 - t_j)$, where $t_j = j / (q + 1)$. In practice, however, we observed that the first few eigenvalues often drop faster, while the decay of the remaining eigenvalues well follows the theory. Empirically, the actual decay resembles an ``S''-shaped curve:
\begin{equation*}
	z_j = z(t_j, \gamma) = \log(1 - t_j) + \gamma\log(1 / t_j),
\end{equation*}
where $\gamma \in [0, 1]$ is a shape parameter. For example, Figure \ref{fig4} illustrates the decay of $\log(\lambda_j)$ against $t_j$ for the examples in the previous section. The fast decay of the first few eigenvalues and the resulting ``S''-shaped curve are most noticeable for $p = 500$. The Figure also plots $z(t_j, \gamma)$ for various $\gamma$ values. When $\gamma = 0$, it is the asymptotic decay; when $\gamma = 1$, it is a symmetric ``S''-shaped curve similar to the quantile function of the logistic distribution. Moreover, it appears that we can tweak $\gamma$ value to make $z_j$ similar to $\log(\lambda_j)$.

\begin{figure}
	\includegraphics[width = \columnwidth]{"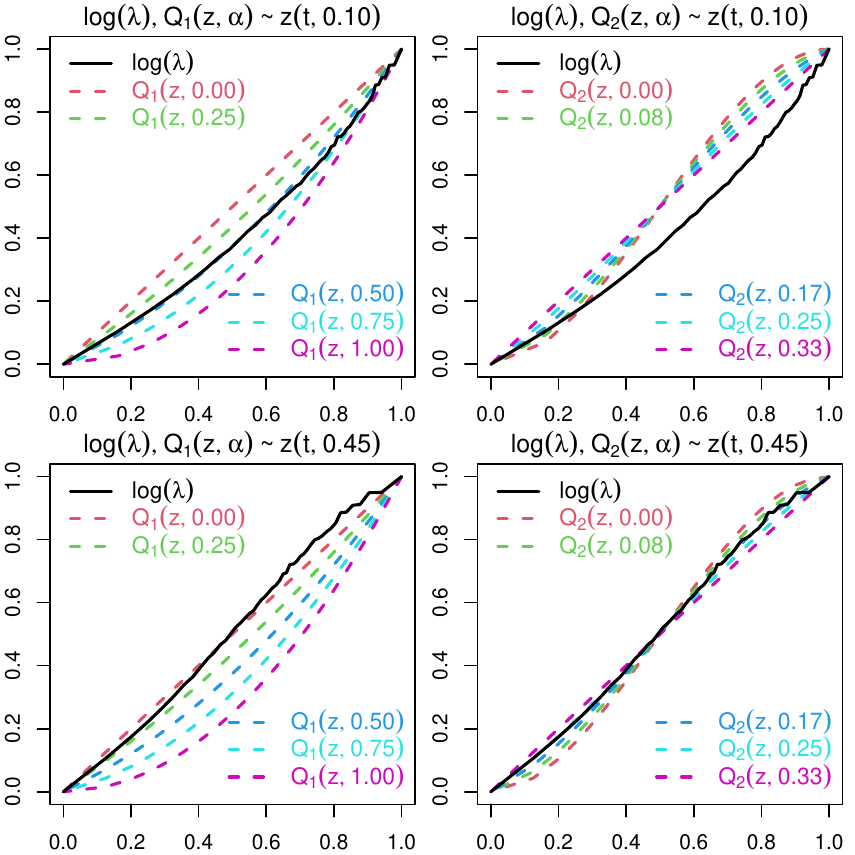"}
	\caption{We can approximate $\log(\lambda)$ by $\log(\hat{\lambda}_j) = Q(z_j,\alpha)$, if $\gamma$ value is properly chosen for $z_j = z(t_j, \gamma)$ such that $\log(\lambda_j)$ (sketched against $z_j$) largely lies on the ``paths'' of $Q(z_j,\alpha)$ as $\alpha$ varies. In this example (the $p = 50$ case in Figure \ref{fig4}), $\gamma = 0.1$ is good for $Q_1(z_j,\alpha)$, but not for $Q_2(z_j,\alpha)$; whereas $\gamma = 0.45$ is good for $Q_2(z_j,\alpha)$, but not for $Q_1(z_j,\alpha)$. (Note: the ranges of $\log(\lambda_j)$ and $z_j$ have been transformed to [0, 1] to aid shape comparison.)}
	\label{fig5}
\end{figure}

As it is generally not possible to tweak $\gamma$ value to make $z_j$ identical to $\log(\lambda_j)$, we chose to model $\log(\lambda_j)$ as a function of $z_j$, i.e., $\log(\hat{\lambda}_j) = Q(z_j)$. For convenience, let $a = \log(\lambda_q)$ and $b = \log(\lambda_1)$. We also transform the range of $z_j$ to $[0, 1]$ by $z_j \leftarrow (z_j - z_q) / (z_1 - z_q)$. As we require $\hat{\lambda}_q = \lambda_q$ and $\hat{\lambda}_1 = \lambda_1$, the function must satisfy $Q(0) = a$ and $Q(1) = b$. To be able to determine $Q(z_j)$ with the last constraint $\sum_{j = 1}^q\hat{\lambda}_j = q\bar{\lambda}$, we can only parameterize the function with one unknown. We now denote this function by $Q(z_j, \alpha)$ and suggest two parametrizations that prove to work well in practice. The first one is a quadratic polynomial:
\begin{equation*}
	Q_1(z_j, \alpha) = a + (b - a)z_j + \alpha(z_j ^ 2 - z_j).
\end{equation*}
It is convex and monotonically increasing when $\alpha \in [0, b - a]$. The second one is a cubic polynomial:
\begin{equation*}
	\begin{split}
	Q_2(z_j,\alpha) =&\ [c_0(z_j) + c_2(z_j)]a\ +\\
	&\ [c_2(z_j) + c_3(z_j)]b\ +\\
	&\ [c_1(z_j) - c_2(z_j)]\alpha,
	\end{split}
\end{equation*}
represented using cubic Bernstein polynomials:
\begin{align*}
	c_0(z_j) &= (1 - z_j) ^ 3, & c_1(z_j) & = 3z_j(1 - z_j)^2,\\
	c_2(z_j) &= 3z_j^2(1 - z_j), & c_3(z_j) &= z_j^3.
\end{align*}
It is an ``S''-shaped curve and if $\alpha \in [a, (2a + b)/3]$, it is monotonically increasing, concave on [0, 0.5] and convex on [0.5, 1]. See Figure \ref{fig5} for what the two functions look like as $\alpha$ varies. To facilitate computation, we rewrite both cases as:
\begin{equation*}
	\log(\hat{\lambda}_j) = Q(z_j,\alpha) = \theta_j + h_j\alpha, \kern 2mm \alpha \in [\alpha_l,\alpha_r].
\end{equation*}
Finding $\alpha$ such that $\sum_{j = 1}^q\hat{\lambda}_j = q\bar{\lambda}$ is equivalent to finding the root of $g(\alpha) = \sum_{j = 1}^q\exp(\theta_j + h_j\alpha) - q\bar{\lambda}$. As $g(\alpha)$ is differentiable, we use Newton's method (see Algorithm \ref{algorithm: root-finding}) for this task. Obviously, a root exists in $[\alpha_l,\alpha_r]$ if and only if $g(\alpha_l)\cdot g(\alpha_r) <= 0$. Therefore, we are not able to obtain approximated eigenvalues $(\hat{\lambda}_j)_1^q$ if this condition does not hold. Intuitively, the condition is met if $\log(\lambda_j)$, when sketched against $z_j$, largely lies on the ``paths'' of $Q(z_j,\alpha)$ as $\alpha$ varies (see Figure \ref{fig5}). This in turn implies that $\gamma$ value should be properly chosen for $z_j = z(t_j, \gamma)$. For example, Figure \ref{fig5} shows that $\gamma = 0.1$ is good for $Q_1(z_j,\alpha)$, but not for $Q_2(z_j,\alpha)$; whereas $\gamma = 0.45$ is good for $Q_2(z_j,\alpha)$, but not for $Q_1(z_j,\alpha)$.

\begin{figure}
	\includegraphics[width = \columnwidth]{"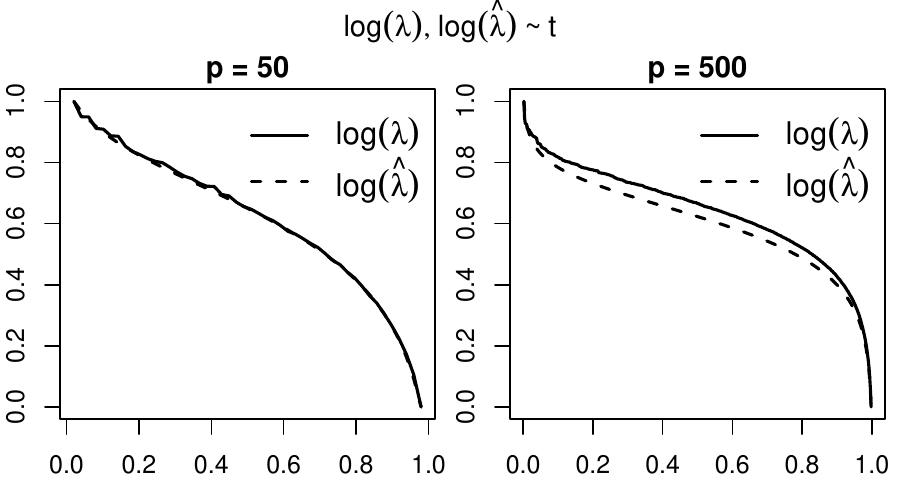"}
	\caption{Log eigenvalues $\log(\lambda_j)$ (solid) and their heuristic approximation $\log(\hat{\lambda}_j)$ (dashed) against $t_j$.}
	\label{fig6}
\end{figure}

\begin{figure}
	\includegraphics[width = \columnwidth]{"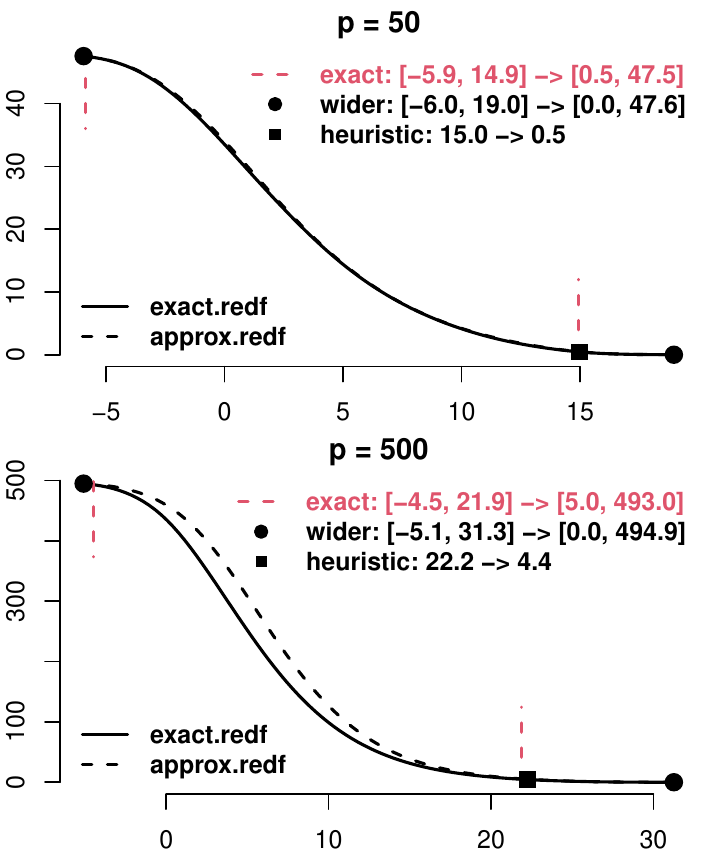"}
	\caption{Approximated $\REDF(\rho)$ (black dashed) and heuristic upper bound $\maxrhohat$ (black square). This Figure enhances Figure \ref{fig2}. As $\maxrhohat$ is closer to $\maxrho$ (red dashed) than $\maxrho^*$ (black circle) is, it is a tighter upper bound.}
	\label{fig7}
\end{figure}

In reality, we don't know what $\gamma$ value is good in advance, so we do a grid search in [0, 1], and for each trial $\gamma$ value, we attempt both $Q_1(z_j,\alpha)$ and $Q_2(z_j,\alpha)$ (see Algorithm \ref{algorithm: ApproxEigen}). We might have several successful approximations, so we take their average for the final $(\hat{\lambda}_j)_1^q$. Figure \ref{fig6} shows $\log(\lambda_j)$ and $\log(\hat{\lambda}_j)$ (both transformed to range between 0 and 1) for the examples demonstrated through Figures \ref{fig2} to \ref{fig5}. The approximation is perfect for small $p$ (which is not surprising because a smaller $p$ means fewer eigenvalues to guess). As $p$ grows, the approximation starts to deviate from the truth, but still remains accurate on both ends (which is important, since the quality of the heuristic upper bound $\maxrhohat$ mainly relies on the approximation to extreme eigenvalues). Figure \ref{fig7} displays (on top of Figure \ref{fig2}) the approximated {\REDF} and the resulting $\maxrhohat$. Clearly, $\maxrhohat$ is closer to $\maxrho$ than $\maxrho^*$ is. Thus, it is a tighter bound. (More simulations of $\maxrho$, $\maxrho^*$ and $\maxrhohat$ are conducted in Section \ref{section: Simulations for heuristic upper bound}.)

\begin{algorithm}
	\caption{Approximate eigenvalues of $\bm{E^{\trans}E}$. Inputs: (i) $q$, number of eigenvalues, (ii) $\lambda_1$, $\lambda_q$ and $\bar{\lambda}$, max/min/mean eigenvalues. Quantities with subscript $j$ are calculated for $j = 1, 2, \ldots, q$. For the inputs of Algorithm \ref{algorithm: root-finding}, use $(\alpha_l + \alpha_r) / 2$ for the initial value and bound the size of a Newton step by $\delta_{\max} = (\alpha_r - \alpha_l) / 4$.}
	\label{algorithm: ApproxEigen}
	\begin{algorithmic}[1]
		\State $a = \log(\lambda_q)$, $b = \log(\lambda_1)$
		\State $t_j = j / (q + 1)$
		\State $\hat{\lambda}_j = 0$ \Comment {initialize approximation}
		\State $N = 0$ \Comment {number of successes}
		\For {$\gamma = 0,\ 0.05,\ 0.10,\ \ldots,\ 1$}
		\State $z_j' = \log(1 - t_j) - \gamma\log(t_j)$ \Comment {try $z(t_j, \gamma)$}
		\State $z_j = (z_j' - z_q') / (z_1' - z_q')$
		\State $\theta_j = a + (b - a)z_j$ \Comment {try $Q_1(z_j, \alpha)$}
		\State $h_j = z_j^2 - z_j$
		\State $\alpha_l = 0$, $\alpha_r = b - a$
		\If {$g(\alpha_l)g(\alpha_r) <= 0$}
		\State find $g(\alpha)$'s root $\alpha$ using Algorithm \ref{algorithm: root-finding}
		\State $N = N + 1$
		\State $\hat{\lambda}_j = \hat{\lambda}_j + \exp(\theta_j + \alpha h_j)$
		\EndIf
		\State $c_{0j} = (1 - z_j)^3$ \Comment {try $Q_2(z_j, \alpha)$}
		\State $c_{1j} = 3z_j(1 - z_j)^2$
		\State $c_{2j} = 3z_j^2(1 - z_j)$
		\State $c_{3j} = z_j^3$
		\State $\theta_j = a(c_{0j} + c_{2j}) + b(c_{2j} + c_{3j})$
		\State $h_j = c_{1j} - c_{2j}$
		\State $\alpha_l = a$, $\alpha_r = (2a + b) / 3$
		\If {$g(\alpha_l)g(\alpha_r) <= 0$}
		\State find $g(\alpha)$'s root $\alpha$ using Algorithm \ref{algorithm: root-finding}
		\State $N = N + 1$
		\State $\hat{\lambda}_j = \hat{\lambda}_j + \exp(\theta_j + \alpha h_j)$
		\EndIf
		\EndFor
		\If {$N > 0$} \Comment {average over successes}
		\State $\hat{\lambda}_j = \hat{\lambda}_j / N$
		\Else
		\State Warning: unable to approximate $\lambda_j$
		\EndIf\\
		\Return $\lambda_j$
	\end{algorithmic}
\end{algorithm}

In summary, to get the heuristic upper bound $\maxrhohat$, we first apply Algorithm \ref{algorithm: ApproxEigen} to compute $(\hat{\lambda}_j)_1^q$ for approximating {\REDF} \eqref{eqn: redf}, then apply Algorithm \ref{algorithm: root-finding} to solve the approximated {\REDF} for $\maxrhohat$. Both steps are computationally efficient at $O(p)$ cost, as they do not involve matrix computations. Thus, the overall computational cost of the heuristically improved interval $[\minrho^*, \maxrhohat]$ remains $O(p^2)$. The computation is also fully automatic. In the first step, all variable inputs required by Algorithm \ref{algorithm: ApproxEigen}, namely $q$, $\lambda_1$, $\lambda_q$ and $\bar{\lambda}$, can be obtained during the computation of the wider interval $[\minrho^*, \maxrho^*]$. In the second step, we can automate Algorithm \ref{algorithm: root-finding} by using $(\minrho^* + \maxrho^*)/2$ for the initial $\rho$ value, and bounding the size of a Newton step by $\delta_{\max} = (\maxrho^* - \minrho^*)/4$.

In rare cases, Algorithm \ref{algorithm: ApproxEigen} fails to approximate eigenvalues. On this occasion, $\maxrhohat$ is not available and we have to stick to $\maxrho^*$. In principle, the chance of failure can be reduced by enhancing the heuristic. For example, we may introduce a shape parameter $\nu \geq 1$ and model the ``S''-shaped decay as $z_j = z(t_j, \gamma, \nu) = \log(1 - t_j) + \gamma[\log(1 / t_j)]^{\nu}$, which allows the first few eigenvalues to decrease even faster. We may also propose $Q(z_j, \alpha)$ of new parametric forms. In short, Algorithm \ref{algorithm: ApproxEigen} can be easily extended for improvement.

\subsection{Summary}

We propose two search intervals for $\rho$: the exact one $[\minrho, \maxrho]$ and the wider one $[\minrho^*, \maxrho^*]$. We prefer the wider one to the exact one for three reasons. Firstly, the wider one has a closed-form formula and its computation is fully automatic, whereas the exact one is implicitly defined by root-finding and its computation is not automatic by itself. Secondly, the wider one is computationally efficient. Its $O(p^2)$ cost is no greater than the $O(Np^2)$ cost for solving PLS over $N$ trial $\rho$ values, whereas the $O(p^3)$ cost for computing the exact one is too expensive. Thirdly, the wider one, or rather, its upper bound $\maxrho^*$, can be tightened using simple heuristics, and the heuristic upper bound $\maxrhohat$ can be computed at only $O(p)$ cost. To reflect the practical impact of computational costs, we report in Table \ref{table: computation time} the actual runtime of different computational tasks for growing $p$.

\begin{table}
	\centering
	\begin{minipage}{\columnwidth}
	\caption{Runtime (in seconds (s) or milliseconds (1ms = 0.001s) of different computational tasks for growing $p$, on an Intel i5-8250U CPU @ 1.60GHz.}
	\label{table: computation time}
	\begin{tabular}{ccccc}
		\toprule
		$p$ & 500 & 1000 & 1500 & 2000\\
		\midrule
		$\minrho^*, \maxrho^*, \maxrhohat$ & \phantom{00}4.74ms & 0.02s & 0.07s & \phantom{0}0.15s\\
		$\minrho, \maxrho$ & 156.50ms & 1.85s & 7.48s & 17.37s\\
		PLS ($N = 20$) & \phantom{0}58.30ms & 0.25s & 0.58s & \phantom{0}1.00s\\
		\botrule
	\end{tabular}
	\end{minipage}
\end{table}

\section{Simulations}
\label{section: Simulations for heuristic upper bound}

The exact interval $[\minrho, \maxrho]$ and the wider one $[\minrho^*, \maxrho^*]$ have known theoretical property. By construction, we have (for $\kappa = 0.01$):
\begin{align*}
	\minrho^* &\le \minrho, & \REDF(\minrho^*) &\ge \REDF(\minrho) = 0.99q\\
	\maxrho^* &\ge \maxrho, & \REDF(\maxrho^*) &\le \REDF(\maxrho) = 0.01q.
\end{align*}
However, the heuristic bound $\maxrhohat$ has unknown property. We expect it to be tighter than $\maxrho^*$, in the sense that it is closer to $\maxrho$. Ideally, it should satisfy $\maxrho \le \maxrhohat \le \maxrho^*$. This is supported by Figure \ref{fig7}, but we still need extensive simulations to be confident about this in general.

For comprehensiveness, let's experiment every possible setup for penalized B-splines that affects $\REDF(\rho)$. The placement of equidistant or unevenly spaced knots produces different $\bm{B}$. The choice of difference or derivative penalty gives different $\bm{D}_m$. We also consider weighted data where $(x_i, y_i)$ has weight $w_i$. This leads to a penalized weighted least squares problem (PWLS) that can be transformed to a PLS problem by absorbing weights into $\bm{B}$: $\bm{B} \leftarrow \bm{W}^{1/2}\bm{B}$, where $\bm{W}$ is a diagonal matrix with element $W_{ii} = w_i$. Altogether, we have 8 scenarios (see Table \ref{table: 8 scenarios}).

\begin{table}
	\centering
	\begin{minipage}{0.7\columnwidth}
	\caption{The 8 scenarios for simulations}
	\label{table: 8 scenarios}
	\begin{tabular}{cccc}
		\toprule
		& \parbox[t][\totalheight][s]{10ex}{derivative\\ \phantom{x}penalty} & \parbox[t][\totalheight][s]{12ex}{equidistant\\ \phantom{xx}knots} & \parbox[t][\totalheight][s]{9ex}{weighted\\ \phantom{xx}data}\\
		\midrule
		1 & FALSE & FALSE & FALSE\\
		2 & TRUE & FALSE & FALSE\\
		3 & FALSE & TRUE & FALSE\\
		4 & TRUE & TRUE & FALSE\\
		\midrule
		5 & FALSE & FALSE & TRUE\\
		6 & TRUE & FALSE & TRUE\\
		7 & FALSE & TRUE & TRUE\\
		8 & TRUE & TRUE & TRUE\\
		\botrule
	\end{tabular}
	\end{minipage}
\end{table}

Here are more low-level details for setting up $\bm{B}$, $\bm{D}_m$ and $\bm{W}$. In general, to construct order-$d$ B-splines $\Bcal_j(x)$, $j = 1, \ldots, p$, we need to place $p + d$ knots $(\xi_k)_1^{p + d}$. For our simulations, we take  $\xi_k = k$ for equidistant knots and $\xi_k \sim \textrm{N}(k, [(p + d)/10]^2)$ for unevenly spaced knots. We then generate 10 uniformly distributed $x$ values between every two nearby knots for constructing the design matrix $\bm{B}$ and the penalty matrix $\bm{D}_m$. For the weights, we use random samples from Beta(3, 3) distribution.

To finish our setup, we still need to specify the B-spline order $d$, the penalty order $m$ and basis dimension $p$. By design, each combination of these parameters has 8 scenarios. For our simulations, we are to test different $(d, m)$ pairs for growing $p$. We call each combination of $d$, $m$, $p$ and scenario ID an experiment.

\begin{figure}
	\includegraphics[width = \columnwidth]{"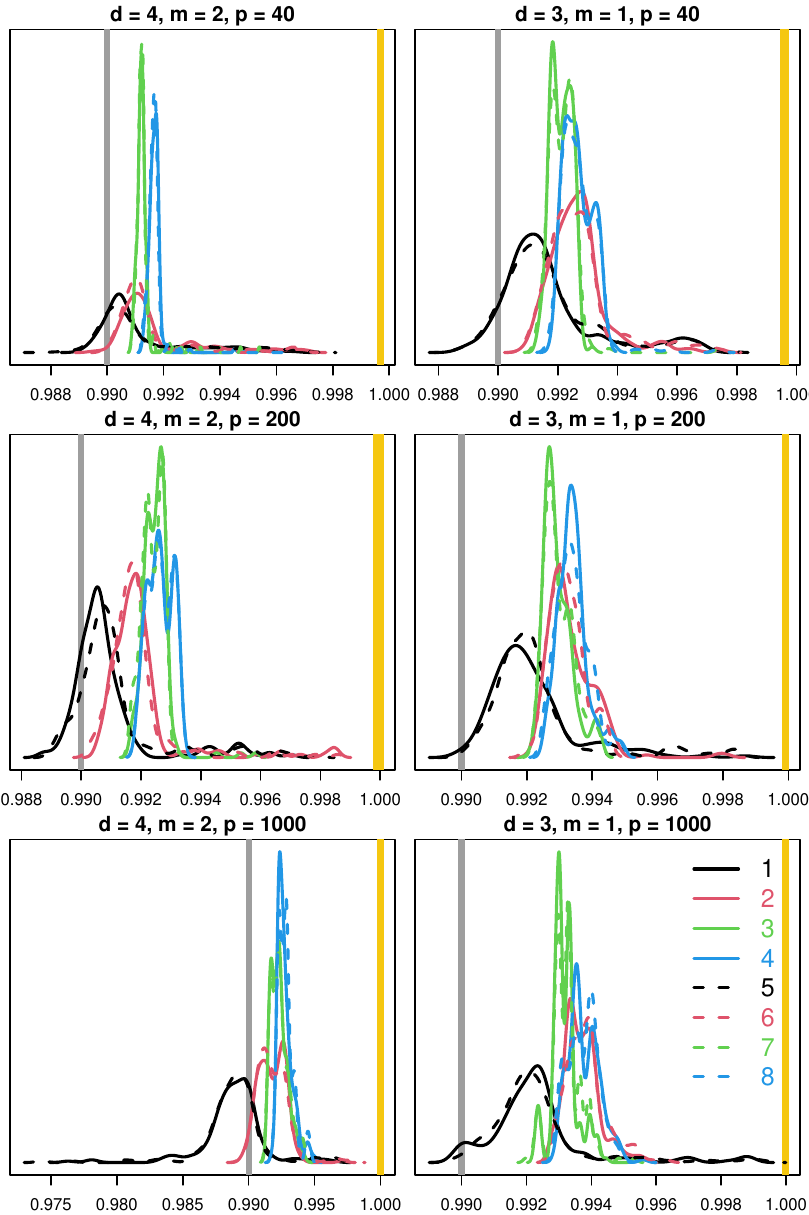"}
	\caption{Estimated probability density function of $\textrm{P}(\maxrhohat)$, i.e., the proportion of $[0, q]$ covered by $[\REDF(\maxrhohat), q]$. If $\maxrho \le \maxrhohat \le \maxrho^*$, the density curve should lie between 0.99 (gray line) and $\textrm{P}(\maxrho^*)$ (yellow line). Numbers 1 to 8 refer to the scenarios in Table \ref{table: 8 scenarios}.}
	\label{fig8}
\end{figure}

In principle, we want to run an experiment, say 200 times, and see how $\maxrhohat$ is distributed relative to $\maxrho$ and $\maxrho^*$. But as none of these quantities stays fixed between runs, visualization is not easy. Back to our \REDF-oriented thinking, let's compare the proportion of $[0, q]$ covered by  $[\REDF(\maxrho), q]$, $[\REDF(\maxrhohat), q]$ and $[\REDF(\maxrho^*), q]$:
\begin{align*}
	\textrm{P}(\maxrho) &= 1 - \REDF(\maxrho) / q = 0.99,\\
	\textrm{P}(\maxrhohat) &= 1 - \REDF(\maxrhohat) / q,\\
	\textrm{P}(\maxrho^*) &= 1 - \REDF(\maxrho^*) / q.
\end{align*}
Since $\REDF(\rho)$ is decreasing, we shall observe $0.99 \le \textrm{P}(\maxrhohat) \le \textrm{P}(\maxrho^*)$ if we expect $\maxrho \le \maxrhohat \le \maxrho^*$. Interestingly, our simulations show that the variance of $\textrm{P}(\maxrho^*)$ is so low that its probability density function almost degenerates to a vertical line through its mean. This implies that we need to check if the density curve of $\textrm{P}(\maxrhohat)$ lies between two vertical lines. Figure \ref{fig8} shows our simulation results for cubic splines with a 2nd order penalty and quadratic splines with a 1st order penalty as $p$ grows. They look satisfying except for scenarios 1 and 5. In these cases, a non-negligible proportion of the density curve breaches $0.99$, so we will get $\maxrhohat < \maxrho$ occasionally. Therefore, the resulting interval $[\minrho^*, \maxrhohat]$ is narrower than $[\minrho, \maxrho]$. However, it is wide enough for grid search, as the corresponding {\REDF} range still covers a significant proportion of $[0, q]$ (check out the numbers labeled along the x-axis of each graph in the Figure). So, empirically, $\maxrhohat$ is a fair tight upper bound.

One anonymous reviewer asked if lowering the value of $\kappa$ would improve the performance of $\maxrhohat$ for scenarios 1 and 5. The answer is no, because $\kappa$ has no effect on the heuristic approximation to eigenvalues (see Algorithm \ref{algorithm: ApproxEigen}). To really improve the performance of $\maxrhohat$, we need to enhance our heuristics rather than alter the value of $\kappa$. Figure S1 in the supplementary material shows further simulation results with $\kappa = 0.005$ and 0.001. The results are very similar to Figure \ref{fig8}, confirming that the value of $\kappa$ is not critical.

\section{Application}
\label{section: application}

As an application of our automatic search interval to practical smoothing, we revisit the COVID-19 data example in the Introduction. To stress that our interval is criterion-independent, we illustrate smoothness selection using both GCV and REML.

\begin{figure}
	\includegraphics[width = \columnwidth]{"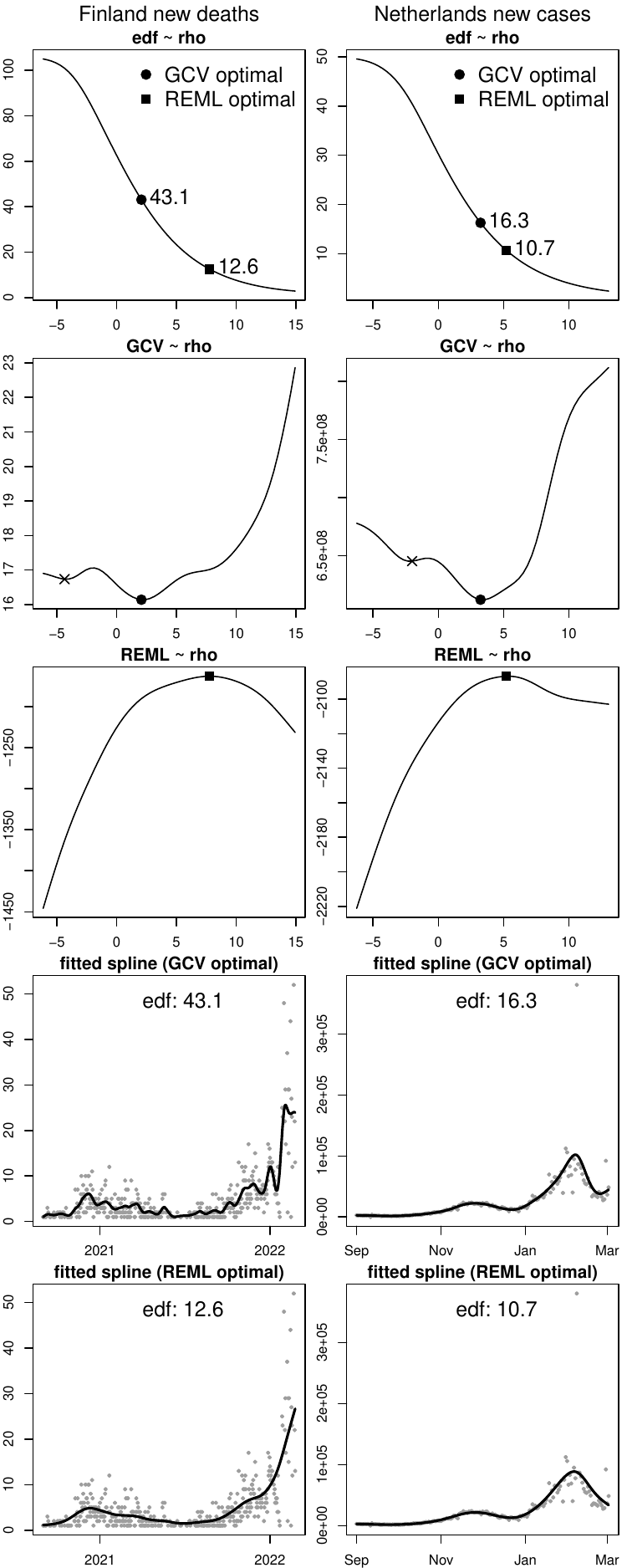"}
	\caption{Smoothing daily COVID-19 data in Finland ($n = 408$ data; displayed in column 1) and Netherlands ($n = 182$ data; displayed in column 2). A cubic general P-splines with $n / 4$ knots and a 2nd order difference penalty is used for smoothing. For both examples, the optimal $\rho$ value chosen by REML is bigger than that selected by GCV, resulting in a smoother yet more plausible fit.}
	\label{fig9}
\end{figure}

The Finland dataset reports new deaths on 408 out of 542 days from 2020-09-01 to 2022-03-01. The Netherlands dataset reports new cases on 181 out of 182 days from 2021-09-01 to 2022-03-01. Let $n$ be the number of data. For smoothing we set up cubic B-splines on $n / 4$ knots placed at equal quantiles of the reporting days, and penalize them by a 2nd order difference penalty matrix. For the Finland example, our computed search interval is $[\minrho^*, \maxrhohat] = [-6.15, 14.93]$ (with $\maxrho^* = 20.25$). For the Netherlands example, the search interval is $[\minrho^*, \maxrhohat] = [-6.26, 13.05]$ (with $\maxrho^* = 16.97$). Figure \ref{fig9} sketches $\EDF(\rho)$, $\textrm{GCV}(\rho)$ and $\textrm{REML}(\rho)$ on the search intervals. For both examples, GCV has a local minimum and a global minimum (see Figure \ref{fig1} for a zoomed-in display), whereas REML has a single maximum. In addition, the optimal $\rho$ value chosen by REML is bigger than that selected by GCV, yielding a smoother yet more plausible fit. In fact, theoretical properties of both criteria have been well studied by \citet{REML-is-better-than-gcv}. In short, GCV is more likely to have multiple local optima. It is also more likely to underestimate the optimal $\rho$ and cause overfitting. Thus, REML is superior to GCV for smoothness selection. But anyway, the focus here is not to discuss the choice of the selection criterion, but to demonstrate that our automatic search interval for $\rho$ is wide enough for exploring any criterion.

\begin{figure}
	\includegraphics[width = \columnwidth]{"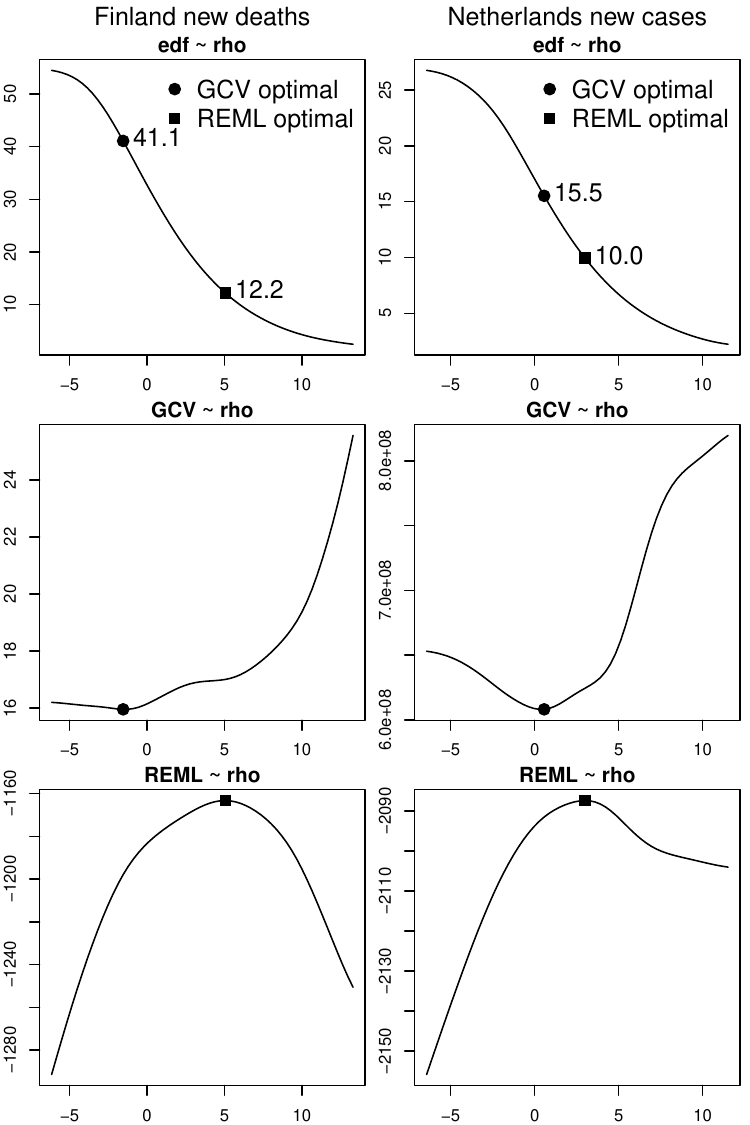"}
	\caption{The altered $\EDF(\rho)$, $\textrm{GCV}(\rho)$ and $\textrm{REML}(\rho)$ after we halve the number of knots in the COVID-19 smoothing example. Interestingly, GCV no longer has a second local minimum, yet it still underestimates the optimal $\rho$ value when compared with REML.}
	\label{fig10}
\end{figure}

The {\EDF} curves in Figure \ref{fig9} show that more than half of the B-spline coefficients are suppressed by the penalty in the optimal fit. In the Finland case, the maximum possible {\EDF} in Figure \ref{fig9} is 106, but the optimal {\EDF} (either 43.1 or 12.6) is less than half of that. To avoid such waste, we can halve the number of knots, i.e., place $n / 8$ knots. This alters the design matrix, the penalty matrix and hence $\EDF(\rho)$, so the search interval needs be recomputed. The new interval is $[\minrho^*, \maxrhohat] = [-6.16, 13.30]$ (with $\maxrho^* = 17.47$) for the Finland example and $[\minrho^*, \maxrhohat] = [-6.43, 11.51]$ (with $\maxrho^* = 14.66$) for the Netherlands example. Figure \ref{fig10} sketches $\EDF(\rho)$, $\textrm{GCV}(\rho)$ and $\textrm{REML}(\rho)$ on their new range. Interestingly, $\textrm{GCV}(\rho)$ no longer has a second local minimum, yet it still underestimates the optimal $\rho$ value when compared with REML. The fitted splines are almost identical to their counterparts in Figure \ref{fig9}, and thus not shown in the Figure.

\section{Discussion}

We have developed algorithms to automatically produce a search interval for the grid search of the smoothing parameter $\rho$ in penalized splines \eqref{eqn: PLS}. Our search interval has four properties. (i) It gives a safe $\rho$ range where the PLS problem is numerically solvable. (ii) It does not depend on the choice of the smoothness selection criterion. (iii) It is wide enough to contain the global optimum of any criterion. (iv) It is computationally cheap compared with the grid search itself.

The Demmler-Reinsch eigenvalues $(\lambda_j)_1^q$ play a pivotal role in our methodology development. They reveal a one-to-one correspondence between $\rho$ (varying in $(-\infty, +\infty)$) and {\REDF} \eqref{eqn: redf} (varying in $[0, q]$), which motivates us to back-transform a target {\REDF} range $[q\kappa,\ q(1-\kappa)]$ for a suitable $\rho$ range $[\minrho,\maxrho]$, where $\kappa$ is a coverage parameter such that the target {\REDF} range covers $100(1 - 2\kappa)$\% of $[0, q]$. As such, the search interval satisfies (i)-(iii) naturally. To achieve (iv), the computational strategy for the interval needs to meet different goals, depending on whether the PLS problem \eqref{eqn: PLS} is dense or sparse.
\begin{itemize}
	\item A dense PLS problem has $O(p^3)$ computational cost. This applies to penalized splines with a dense design matrix $\bm{B}$ and/or a dense penalty matrix $\bm{S}$. Examples are  truncated power basis splines \citep{SemiPar-book-2003}, natural cubic splines \citep[section 5.3.1]{Wood-GAMs-book} and thin-plate splines \citep[section 5.5.1]{Wood-GAMs-book};
	\item A sparse PLS problem has $O(p^2)$ computational cost. This applies to penalized B-splines (for which the PLS objective is \eqref{eqn: penalized B-splines}) with a sparse design matrix $\bm{B}$ and a sparse penalty matrix $\bm{D}_m$. Examples are O-splines \citep{O-splines-1}, standard P-splines \citep{P-splines} and general P-splines \citep{gps-paper}.
\end{itemize}
Therefore, the computational costs of our search intervals need be no greater than $O(p^3)$ and $O(p^2)$, respectively.

We have made great efforts to optimize our computational strategy for the sparse penalized B-splines. The key is to give up computing all the eigenvalues (as an eigendecomposition has $O(p^3)$ cost) and compute only the mean, the maximum and the minimum eigenvalues. This allows us to obtain a wider interval $[\minrho^*, \maxrho^*] \supseteq [\minrho, \maxrho]$ at $O(p^2)$ cost. We can further tighten this interval using some heuristics. Precisely, step by step, we compute:
\begin{enumerate}
	\item $\bar{\lambda}$, using \eqref{eqn: MeanEigen};
	\item $\lambda_1$ and $\lambda_q$, using Algorithms \ref{algorithm: MaxEigen} and \ref{algorithm: MinEigen};
	\item $[\minrho^*, \maxrho^*]$, using \eqref{eqn: rho bounds};
	\item $(\hat{\lambda}_j)_1^q$, using Algorithm \ref{algorithm: ApproxEigen};
	\item $\maxrhohat$, by applying Algorithm \ref{algorithm: root-finding} to \eqref{eqn: redf} (with $\lambda_j$ replaced by $\hat{\lambda}_j$), \eqref{eqn: redf bounds} and \eqref{eqn: exact rho bounds}. (To automatically start the algorithm, use $(\minrho^* + \maxrho^*) / 2$ for the initial value and set $\delta_{\max} = (\maxrho^* - \minrho^*) / 4$.)
\end{enumerate}
Steps 1-2 each have $O(p^2)$ cost; step 3 is simple arithmetic; steps 4-5 each have $O(p)$ cost. They are implemented by function \texttt{gps2GS} in our \textbf{R} package \textbf{gps} ($\ge$ version 1.1). The function solves the PLS problem \eqref{eqn: penalized B-splines} for a grid of $\rho$ values in $[\minrho^*, \maxrhohat]$. It can be embedded in more advanced statistical modeling methods that rely on penalized splines, such as robust smoothing and generalized additive models.

One anonymous reviewer commented that it appears possible to bypass the use of eigenvalues. It was noticed that the initial $\EDF$ formula \eqref{eqn: edf-via-pls} is free of eigenvalues. So, it was suggested that we could work with $\REDF = \EDF - m = \|\bm{K}^{\inv}\bm{L}\|_F^2 - m$ instead of \eqref{eqn: redf}, when back-transforming a target {\REDF} range for $[\minrho, \maxrho]$ via root-finding. Since the computational complexity of \eqref{eqn: edf-via-pls} is $O(p^2)$, and a root-finding algorithm only takes a few iterations to converge, this method should have $O(p^2)$ cost, too. While this is true, it is not practicable unless we automate the root-finding. If we use Newton's method (see Algorithm \ref{algorithm: root-finding}) for this task, we need an initial value and a maximum stepsize for $\rho$; if we use bisection or Brent's method, we need a search interval for $\rho$. Either way seems to be a deadlock, as we have to supply something relevant to what we hope to find. In fact, whether we express {\REDF} using eigenvalues or not, we will face this difficulty as long as the search interval is implicitly defined by root-finding. This difficulty is not eliminated, until we have a wider interval $[\minrho^*, \maxrho^*]$ in closed form. This interval can be directly computed by \eqref{eqn: rho bounds} using eigenvalues. It can further be exploited to automate the root-finding (for example, see step 5 of our method). In summary, the wider interval and hence the Demmler-Reinsch eigenvalues are key to the automaticity of our method. There is no way to bypass the use of those eigenvalues. The best we can do, is to only compute the maximum, the minimum and the mean eigenvalues required for computing the wider interval.

Since we have to use eigenvalues anyway, we express $\REDF(\rho)$ as \eqref{eqn: redf} (the eigen-form) instead of $\|\bm{K}^{\inv}\bm{L}\|_F^2 - m$ (the PLS-form) when presenting our method. In addition, we prefer the eigen-form to the PLS-form for its simplicity. Given $(\lambda_j)_1^q$, it is easy to compute {\REDF} and its derivative if using the eigen-form, and Newton's method has $O(p)$ cost. By contrast, working with the PLS-form requires matrix computations and matrix calculus so that Newton's method has $O(p^2)$ cost. The difficulty with the eigen-form is the $O(p^3)$ computational cost behind $(\lambda_j)_1^q$. However, once we replace them by their heuristic approximation $(\hat{\lambda}_j)_1^q$ that can be obtained at $O(p)$ cost, the simplicity of the eigen-form becomes real computational efficiency.

It may still be asked why we would rather do some heuristic approximation to tighten our wider interval (as steps 4-5 of our method show), than compute the exact interval by applying Algorithm \ref{algorithm: root-finding} to the PLS-form, \eqref{eqn: redf bounds} and \eqref{eqn: exact rho bounds} (where we can use the wider interval to automate the root-finding). After all, the $O(p^2)$ cost behind the PLS-form, while greater the $O(p)$ cost behind the eigen-form, is no greater than the $O(p^2)$ cost for computing the wider interval. Thus, the overall computational cost of our method would still be $O(p^2)$. While this is true, in our view, it is awkward to compute the exact interval; or at least, it is not worth it. The PLS-form is coupled with PLS solving. If we apply this idea, we would have to do PLS solving on three different sets of $\rho$ grids: one for computing $\minrho$, one for $\maxrho$ and one for the grid search on $[\minrho, \maxrho]$. The first two rounds of PLS solving are a waste, given that we are only interested in the final grid search for smoothness selection. We want to separate the computation of our search interval from PLS solving and grid search, and thus deprecate this idea.

Another comment from the same reviewer, is that we have restricted the search for the optimal $\rho$ in our bounded search interval, so that $\rho = -\infty$ or $+\infty$ can not be chosen. This is a good point. In terms of our {\REDF}-oriented thinking, it means that $\REDF = 0$ and $\REDF = q$ have been excluded. This is deliberately done. By construction, the minimum and the maximum $\REDF$ we can reach are $q\kappa$ and $q(1 - \kappa)$, respectively. Apparently, if we want to approximately cover these two endpoints, we can choose a very small $\kappa$ value. But this is not a good strategy (see the next paragraph for a discussion on the choice of $\kappa$). Instead, we compute the GCV error and the REML score for these limiting cases separately, then include them in our grid search.  In either case, the PLS problem \eqref{eqn: penalized B-splines} degenerates to an ordinary least squares (OLS) problem. Thus, we can work with these OLS problems instead for these quantities. See Appendix \ref{Appendix C} for details.

In this paper, the coverage parameter $\kappa$ is fixed at 0.01. Although reducing this value would widen the target {\REDF} range, we don't recommend it. In reality, $\REDF(\rho)$ quickly plateaus as it gets close to either endpoint (for example, see Figure \ref{fig2}). \kern0pt Setting $\kappa$ too small will result in long flat ``tails'' on both sides. Moreover, $\rho$ values on each ``tail'' give similar $\REDF$ values, fitted splines, GCV errors and REML scores. This is undesirable for grid search. When a fixed number of equidistant grid points are positioned, the smaller $\kappa$ is, the more trial $\rho$ values fall on those ``tails'', and thus, the fewer trial $\rho$ values are available for exploring the ``ramp'' where $\REDF(\rho)$ varies fast. The $\kappa$ is to our method as the convergence tolerance is to an iterative algorithm. It needs to be small, but not unnecessarily small.

Our method has a limitation: it does not apply to a penalized spline whose design matrix $\bm{B}$ does not have full column rank. This is because we need matrix $\bm{L}$, the Cholesky factor of $\bm{B}^{\trans}\bm{B}$, to derive the eigen-form of \REDF. While it is common to have a full rank design matrix in practical smoothing, a rank-deficient design matrix is not problematic at all. A typical example is when $p > n$, i.e., there are more basis functions than (unique) $x$ values. As \cite{P-splines-book} put it, ``It is impossible to have too many B-splines'' (p. 15). Therefore, how to automatically produce a search interval for $\rho$ in this case remains an interesting yet challenging question.

For dense penalized splines, it is acceptable to compute a full eigendecomposition for all the eigenvalues, because solving the PLS problem has $O(p^3)$ cost anyway. It is not the decomposition of $\bm{EE}^{\trans}$, though, as the PLS objective reverts to \eqref{eqn: PLS} and neither $\bm{D}_m$ nor $\bm{E}$ is defined. So, we need to eliminate $\bm{D}_m$ and $\bm{E}$, and express the matrix using $\bm{S}$. First, \eqref{eqn: matrix E} implies $\bm{EE}^{\trans} = \bm{L}^{\inv}\bm{D}_m^{\trans}\bm{D}_m(\bm{L}^{\inv})^{\trans}$. Then, replacing $\bm{D}_m^{\trans}\bm{D}_m$ by $\bm{S}$ (which is the link between \eqref{eqn: PLS} and \eqref{eqn: penalized B-splines}) gives $\bm{EE}^{\trans} = \bm{L}^{\inv}\bm{S}(\bm{L}^{\inv})^{\trans}$. Therefore, $(\lambda_j)_1^q$ are the positive eigenvalues of $\bm{L}^{\inv}\bm{S}(\bm{L}^{\inv})^{\trans}$. Our goal is to solve {\REDF} \eqref{eqn: redf} for the exact interval, but to automate the root-finding, we need the wider interval first. Precisely, step by step, we compute:
\begin{enumerate}
	\item $(\lambda_j)_1^q$, by computing a full eigendecomposition;
	\item $[\minrho^*, \maxrho^*]$, using \eqref{eqn: rho bounds};
	\item $[\minrho, \maxrho]$, by applying Algorithm \ref{algorithm: root-finding} to \eqref{eqn: redf}, \eqref{eqn: redf bounds} and \eqref{eqn: exact rho bounds}. (To automatically start the algorithm, use $(\minrho^* + \maxrho^*) / 2$ for the initial value and set $\delta_{\max} = (\maxrho^* - \minrho^*) / 4$.)
\end{enumerate}
We didn't seek to optimize our method for dense penalize splines, so there is likely plenty of room for improvement.

\backmatter

\bmhead{Supplementary information} Simulations in Section \ref{section: Simulations for heuristic upper bound} were conducted with $\kappa = 0.01$. The value of $\kappa$ is not critical. The supplementary file contains simulation results with $\kappa = 0.005$ and 0.001, which are very similar to Figure \ref{fig8}.

\bmhead{Acknowledgments}

We thank the Editor, the Associate Editor and two anonymous reviewers for their careful review and their valuable comments. These comments are very helpful for us to improve our work.

\section*{Declarations}

\bmhead{Funding}

Zheyuan Li was supported by the National Natural Science Foundation of China under the Young Scientists Fund (No. 12001166). Jiguo Cao was supported by the Natural Sciences and Engineering Research Council of Canada under the Discovery Grant (RGPIN-2018-06008).

\bmhead{Conflict of interest}

The authors declare that they have no conflict of interest.

\bmhead{Code availability}

\textbf{R} code is on the internet at \url{https://github.com/ZheyuanLi/gps-vignettes/blob/main/gps2.pdf}.

\bmhead{Authors' contributions} Zheyuan Li: \textsl{Conceptualization; Methodology; Software; Validation; Formal Analysis; Writing - Original Draft}. Jiguo Cao: \textsl{Writing - Review \& Editing}.

\begin{appendices}

\section{Penalty Matrix}
\label{Appendix A}

The penalized B-splines family includes O-splines (OS), standard P-splines (SPS) and general P-splines (GPS). They differ in (i) the type of knots for constructing B-splines; (ii) the penalty matrix $\bm{D}_m$ applied to B-spline coefficients. See Table \ref{table: penalized B-splines family} for an overview.

\begin{table}
	\centering
	\begin{minipage}{0.6\columnwidth}
		\caption{Penalized B-splines}
		\label{table: penalized B-splines family}
		\begin{tabular}{ccc}
			\toprule
			& $\bm{D}_m^{\textrm{dif}}$ & $\bm{D}_m^{\textrm{der}}$\\
			\midrule
			UBS & SPS, GPS & OS\\
			NUBS & GPS & OS\\
			\botrule
		\end{tabular}
	\end{minipage}
\end{table}

\begin{figure}
	\centering
	\includegraphics[width = \columnwidth]{"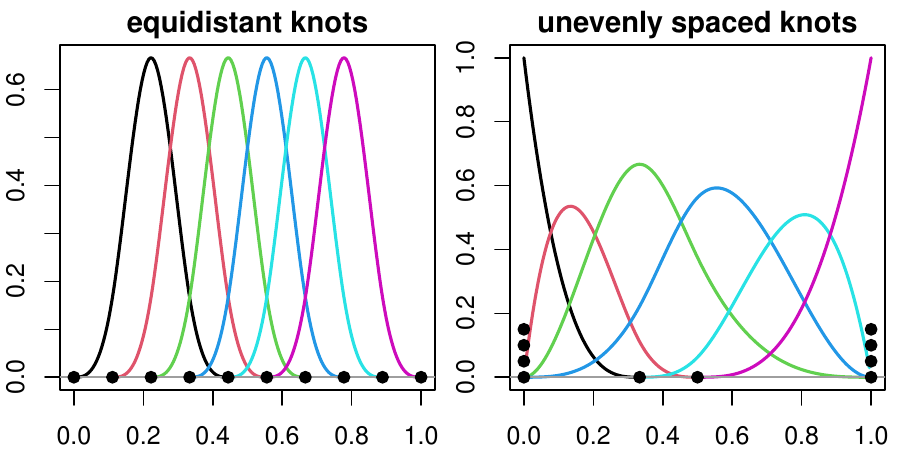"}
	\caption{6 Cubic B-splines $\Bcal_1(x)$, $\Bcal_2(x)$, ... $\Bcal_6(x)$ on 10 knots in [0, 1]. Left: UBS on equidistant knots 0, 1/9, 2/9, ..., 1; Right: NUBS on unevenly spaced knots 0, 0, 0, 0, 1/3, 1/2, 1, 1, 1, 1.}
	\label{figA}
\end{figure}

Knots can be evenly or unevenly spaced. B-splines on equidistant knots are uniform B-splines (UBS); they have identical shapes. B-splines on unevenly spaced knots are non-uniform B-splines (NUBS); they have different shapes. See Figure \ref{figA} for an illustration.

The matrix $\bm{D}_m$ may come from a difference penalty or a derivative penalty. The exact values of its elements also depend on the knot spacing. For details about its derivation and computation, see \citet{gps-paper}. Here, we simply write out some example $\bm{D}_m$ matrices (calculated using function \texttt{SparseD} from our \textbf{R} package \textbf{gps}). For the cubic UBS in Figure \ref{figA}, the 2nd order penalty matrices are:
\begin{gather*}
	\bm{D}_2^{\textrm{dif}} = \left[\begin{array}{rrrrrr}
		1 & -2 & 1\\
		& 1 & -2 & 1\\
		& & 1 & -2 & 1\\
		& & & 1 & -2 & \phantom{+}1
	\end{array}\right],\\
	\bm{D}_2^{\textrm{der}} = \left[\begin{array}{rrrrrr}
		0.19 & -0.29 & 0.00 & 0.10\\
		& 0.25 & -0.44 & 0.11 & 0.07\\
		& & 0.26 & -0.45 & 0.12 & \phantom{\textrm{-}}0.07\\
		& & & 0.18 & -0.36 & \phantom{\textrm{-}}0.18
	\end{array}\right].
\end{gather*}
For the cubic NUBS in Figure \ref{figA}, the 2nd order penalty matrices are:
\begin{gather*}
	\bm{D}_2^{\textrm{dif}} = \left[\begin{array}{rrrrrr}
		54 & -90 & 36\\
		& 24 & -36 & 12.0\\
		& & 9 & -22.5 & 13.5\\
		& & & 18.0 & -42.0 & \phantom{+}24
	\end{array}\right],\\
	\bm{D}_2^{\textrm{der}} = \left[\begin{array}{rrrrrr}
		18 & -26.00 & 6.00 & 2.00\\   
		& 8.94 & -12.75 & 2.80 & 1.01\\   
		& & 4.19 & -7.25 & -1.24 & \phantom{\textrm{-}}4.30\\
		& & & 6.60 & -15.41 & \phantom{\textrm{-}}8.81
	\end{array}\right].
\end{gather*}
Numbers are rounded to 2 decimal places in these matrices.

\section{REML Score}
\label{Appendix B}

The estimation of penalized splines falls within the empirical Bayes framework. Hence, the restricted maximum likelihood (REML) can be used to select the smoothing parameter $\rho$ in penalized splines. This section gives details about the derivation and the computation of REML. For convenience, let's denote the PLS objective \eqref{eqn: penalized B-splines} by $\textrm{PLS}(\bm{\beta})$.

In the Bayesian view, the least squares term in $\textrm{PLS}(\bm{\beta})$ corresponds to Gaussian likelihood:
\begin{equation*}
	\textrm{Pr}(\bm{y}\vert\bm{\beta}) = c_1\cdot
	\exp\left\{-\tfrac{\|\bm{y} - \bm{B\beta}\|^2}{2\sigma^2}\right\},
\end{equation*}
where $c_1 = (2\pi\sigma^2)^{-n/2}$. The wiggliness penalty corresponds to a Gaussian prior:
\begin{equation*}
	\textrm{Pr}(\bm{\beta}) = c_2\cdot
	\exp\left\{-\tfrac{\textrm{e}^{\rho}\|\bm{D}_m\bm{\beta}\|^2}{2\sigma^2}\right\},
\end{equation*}
where $c_2 = (2\pi\sigma^2)^{-(p - m)/2}\cdot\vert\textrm{e}^{\rho}\bm{D}_m\bm{D}_m^{\trans}\vert^{1/2}$ and $\vert\bm{X}\vert$ is the determinant of $\bm{X}$. The unnormalized posterior is then:
\begin{equation*}
	\pi(\bm{\beta}\vert\bm{y})
	= \textrm{Pr}(\bm{y}\vert\bm{\beta}) \cdot \textrm{Pr}(\bm{\beta})
	= c_1 \cdot c_2 \cdot
	\exp\left\{-\tfrac{\textrm{PLS}(\bm{\beta})}{2\sigma^2}\right\}.
\end{equation*}
Clearly, the PLS solution that minimizes $\textrm{PLS}(\bm{\beta})$ is also the posterior mode that maximizes $\pi(\bm{\beta}\vert\bm{y})$. Section \ref{section: penalized B-splines} has shown that the PLS solution is $\bm{\hat{\beta}} = \bm{C}^{\inv}\bm{B^{\trans}y}$, where $\bm{C} = \bm{B^{\trans}B} + \textrm{e}^{\rho}\bm{D}_m^{\trans}\bm{D}_m$. In fact, the Taylor expansion of $\textrm{PLS}(\bm{\beta})$ at $\bm{\hat{\beta}}$ is exactly:
\begin{equation*}
	\textrm{PLS}(\bm{\beta}) = \textrm{PLS}(\bm{\hat{\beta}}) + (\bm{\beta} - \bm{\hat{\beta}})^{\trans}\bm{C}(\bm{\beta} - \bm{\hat{\beta}}).
\end{equation*}
Plugging this into $\pi(\bm{\beta}\vert\bm{y})$ gives:
\begin{equation*}
	\pi(\bm{\beta}\vert\bm{y})
	=\ c_1 \cdot c_2 \cdot c_3 \cdot
	\exp\left\{-\tfrac{(\bm{\beta} - \bm{\hat{\beta}})^{\trans}\bm{C}(\bm{\beta} - \bm{\hat{\beta}})}{2\sigma^2}\right\},
\end{equation*}
where
\begin{equation*}
	c_3 = \exp\left\{-\tfrac{\textrm{PLS}(\bm{\hat{\beta}})}{2\sigma^2}\right\} = \exp\left\{-\tfrac{\textrm{RSS} + \textrm{e}^{\rho}\|\bm{D}_m\bm{\hat{\beta}}\|^2}{2\hat{\sigma}^2}\right\}.
\end{equation*}
The restricted likelihood of $\rho$ and $\sigma^2$ is defined by integrating out $\bm{\beta}$ in $\pi(\bm{\beta}\vert\bm{y})$:
\begin{equation*}
	L_r(\rho, \sigma^2) = \int \pi(\bm{\beta}\vert\bm{y}) \ \textrm{d}\bm{\beta}
	= c_1 \cdot c_2 \cdot c_3 \cdot c_4,
\end{equation*}
where the key integral is:
\begin{equation*}
	\begin{split}
		c_4 &= \int\exp\left\{-\tfrac{(\bm{\beta} - \bm{\hat{\beta}})^{\trans}\bm{C}(\bm{\beta} - \bm{\hat{\beta}})}{2\sigma^2}\right\} \textrm{d}\bm{\beta}\\
		&= (2\pi\sigma^2)^{p/2} \cdot \vert\bm{C}\vert^{-1/2}.
	\end{split}
\end{equation*}
Thus, the restricted log-likelihood, or the REML criterion function, is:
\begin{equation*}
	\begin{split}
		l_r(\rho, \sigma^2) =&\ \log[L_r(\rho, \sigma^2)] = \log(c_1 \cdot c_2 \cdot c_3 \cdot c_4)\\
		=&\ \tfrac{1}{2}\log\vert\textrm{e}^{\rho}\bm{D}_m\bm{D}_m^{\trans}\vert - \tfrac{1}{2}\log\vert\bm{C}\vert\ -\\
		&\ \tfrac{n - m}{2}\log(2\pi\sigma^2) - \tfrac{\textrm{RSS}}{2\sigma^2} - \tfrac{\textrm{e}^{\rho}\|\bm{D}_m\bm{\hat{\beta}}\|^2}{2\hat{\sigma}^2}.
	\end{split}
\end{equation*}
For practical convenience, we can replace $\sigma^2$ by its Pearson estimate $\hat{\sigma}^2 = \tfrac{\textrm{RSS}}{n - \EDF}$. This simplifies the restricted log-likelihood to a function of $\rho$ only:
\begin{equation*}
	\begin{split}
		\textrm{REML}(\rho)
		=&\ \tfrac{1}{2}\log\vert\textrm{e}^{\rho}\bm{D}_m\bm{D}_m^{\trans}\vert - \tfrac{1}{2}\log\vert\bm{C}\vert\ -\\
		&\ \tfrac{n - m}{2}\log(2\pi\hat{\sigma}^2) - \tfrac{n - \EDF}{2} - \tfrac{\textrm{e}^{\rho}\|\bm{D}_m\bm{\hat{\beta}}\|^2}{2\hat{\sigma}^2}.
	\end{split}
\end{equation*}
This is the REML criterion used in this paper and our implementation.

The log-determinants in the REML score can be computed using Cholesky factors. During the computation of $\bm{\hat{\beta}}$ (see Section \ref{section: penalized B-splines} for details), we already obtained the Cholesky factorization $\bm{C} = \bm{KK}^{\trans}$. Therefore,
\begin{gather*}
	\vert\bm{C}\vert = \vert\bm{K}\vert^2 = \textstyle\prod_{j = 1}^{p}K_{jj}^2,\\
	\log\vert\bm{C}\vert = \textstyle2\sum_{j = 1}^{p}\log(K_{jj}),
\end{gather*}
where $K_{jj}$ is the $j$\textsuperscript{th} diagonal element of $\bm{K}$. For the other log-determinant, we first pre-compute the lower triangular Cholesky factor of $\bm{D}_m\bm{D}_m^{\trans}$, denoted by $\bm{F}$. (Thanks to the band sparsity of $\bm{D}_m$, both $\bm{D}_m\bm{D}_m^{\trans}$ and its Cholesky factorization can be computed at $O(p^2)$ cost.) Then for any trial $\rho$ value, we have:
\begin{gather*}
	\vert\textrm{e}^{\rho}\bm{D}_m\bm{D}_m^{\trans}\vert = \textrm{e}^{(p - m)\rho}\vert\bm{D}_m\bm{D}_m^{\trans}\vert = \textrm{e}^{(p - m)\rho}\vert\bm{F}\vert^2,\\
	\log\vert\textrm{e}^{\rho}\bm{D}_m\bm{D}_m^{\trans}\vert = (p - m)\rho + \textstyle2\sum_{j = 1}^{p - m}\log(F_{jj}),
\end{gather*}
where $F_{jj}$ is the $j$\textsuperscript{th} diagonal element of $\bm{F}$.

In summary, quantities in $\textrm{REML}(\rho)$ can be either pre-computed or obtained while computing $\bm{\hat{\beta}}$, $\textrm{RSS}(\rho)$, $\EDF(\rho)$ and $\textrm{GCV}(\rho)$. As a result, for any trial $\rho$ value on a search grid, its REML score can be efficiently computed at $O(p)$ cost.

\section{When \texorpdfstring{$\rho = \pm\infty$}{rho = +/- infty}}
\label{Appendix C}

When $\rho = -\infty$ or $\rho = +\infty$, the PLS problem \eqref{eqn: penalized B-splines} degenerates to an ordinary least squares (OLS) problem.
\begin{itemize}
	\item When $\rho = -\infty$, the penalty term in the PLS objective \eqref{eqn: penalized B-splines} vanishes, leaving only $\|\bm{y} - \bm{B\beta}\|^2$.
	\item When $\rho = +\infty$, the PLS objective \eqref{eqn: penalized B-splines} is $+\infty$ anywhere except when $\bm{D}_m\bm{\beta} = \bm{0}$, i.e., $\bm{\beta}$ is in $\bm{D}_m$'s null space. Let $\bm{N}$ be a $p \times m$ matrix whose columns form an orthonormal basis of this null space. Then, we can write $\bm{\beta} = \bm{N\alpha}$ in terms of a new coefficient vector $\bm{\alpha}$. Thus, the objective becomes $\|\bm{y} - \bm{BN\alpha}\|^2$.
\end{itemize}
Without loss of generality, let's express an OLS objective as $\|\bm{y} - \bm{Xb}\|^2$. Clearly, to match the objective for $\rho = -\infty$, we let $\bm{X} = \bm{B}$ and $\bm{b} = \bm{\beta}$; to match the objective for $\rho = +\infty$, we let $\bm{X} = \bm{BN}$ and $\bm{b} = \bm{\alpha}$.

A limiting PLS problem and its corresponding OLS problem have the same {\EDF}, residuals, fitted values, GCV errors and REML scores. Since the number of coefficients in an OLS problem defines the {\EDF} of the problem, we have $\EDF = p$ for $\rho = -\infty$ and $\EDF = m$ for $\rho = +\infty$. The GCV error can still be computed by $n\cdot\textrm{RSS}/(n - \EDF)^2$, where $\textrm{RSS} = \|\bm{y} - \bm{X\hat{b}}\|^2$ and $\bm{\hat{b}} = (\bm{X}^{\trans}\bm{X})^{\inv}\bm{X}^{\trans}\bm{y}$ is the OLS solution. To derive the REML score, we start with the likelihood for the OLS problem:
\begin{equation*}
	\textrm{Pr}(\bm{y} \vert \bm{b}) = c_1\cdot\exp\left\{-\tfrac{\|\bm{y} - \bm{Xb}\|^2}{2\sigma^2}\right\},
\end{equation*}
where $c_1 = (2\pi\sigma^2)^{-n/2}$. The Taylor expansion of the least squares at $\bm{\hat{b}}$ is exactly:
\begin{equation*}
	\|\bm{y} - \bm{Xb}\|^2 = \textrm{RSS} + (\bm{b} - \bm{\hat{b}})^{\trans}(\bm{X}^{\trans}\bm{X})(\bm{b} - \bm{\hat{b}}).
\end{equation*}
Plugging this into $\textrm{Pr}(\bm{y} \vert \bm{b})$ gives:
\begin{equation*}
	\textrm{Pr}(\bm{y} \vert \bm{b}) = c_1\cdot c_2\cdot\exp\left\{-\tfrac{(\bm{b} - \bm{\hat{b}})^{\trans}(\bm{X}^{\trans}\bm{X})(\bm{b} - \bm{\hat{b}})}{2\sigma^2}\right\},
\end{equation*}
where $c_2 = \exp\{-\tfrac{\textrm{RSS}}{2\sigma^2}\}$. The restricted likelihood of $\sigma^2$ is defined by integrating out $\bm{b}$ in $\textrm{Pr}(\bm{y} \vert \bm{b})$:
\begin{equation*}
	L_r(\sigma^2) = \int \textrm{Pr}(\bm{y} \vert \bm{b})\ \textrm{d}\bm{b} = c_1 \cdot c_2 \cdot c_3,
\end{equation*}
where the key integral is:
\begin{equation*}
	\begin{split}
		c_3 &= \int \exp\left\{-\tfrac{(\bm{b} - \bm{\hat{b}})^{\trans}(\bm{X}^{\trans}\bm{X})(\bm{b} - \bm{\hat{b}})}{2\sigma^2}\right\}\ \textrm{d}\bm{b}\\
		&= (2\pi\sigma^2)^{\EDF/2} \cdot \vert\bm{X}^{\trans}\bm{X}\vert^{-1/2}.
	\end{split}
\end{equation*}
Thus, the restricted log-likelihood is:
\begin{equation*}
	\begin{split}
		l_r(\sigma^2) &= \log[L_r(\sigma^2)] = \log(c_1 \cdot c_2 \cdot c_3)\\
		&= -\tfrac{n - \EDF}{2}\log(2\pi\sigma^2) - \tfrac{1}{2}\log\vert\bm{X}^{\trans}\bm{X}\vert - \tfrac{\textrm{RSS}}{2\sigma^2}.
	\end{split}
\end{equation*}
Replacing $\sigma^2$ by its Pearson estimate $\hat{\sigma}^2 = \tfrac{\textrm{RSS}}{n - \EDF}$ gives the REML score:
\begin{equation*}
	-\tfrac{n - \EDF}{2}[1 + \log(2\pi\hat{\sigma}^2)] - \tfrac{1}{2}\log\vert\bm{X}^{\trans}\bm{X}\vert.
\end{equation*}

\end{appendices}

\end{document}